\documentclass[aps,prd,notitlepage,showpacs,nofootinbib,superscriptaddress]{revtex4-1}
 \pdfoutput=1
\usepackage[usenames,dvipsnames]{color}  
\usepackage{graphicx}
\usepackage{setspace}
\usepackage{multirow}
\usepackage{caption}
\usepackage{amsmath}
\usepackage{slashed}
\usepackage{amssymb}
\usepackage[colorlinks=true,citecolor=darkred,urlcolor=darkred, pdfborder={0 0 0}]{hyperref}
\usepackage[normalem]{ulem}
\usepackage{xcolor}
\usepackage{multirow}
\usepackage{gensymb}
\usepackage{rotating}
\usepackage{braket}
\usepackage{array}
\usepackage{comment}
\usepackage{orcidlink}

\newcolumntype{M}[1]{>{\centering\arraybackslash}m{#1}}

\makeatletter
\def\p@subsection{}
\makeatother

\definecolor{darkred}{rgb}{0.6,0,0}

\definecolor{linkcolor}{rgb}{0,0,0.5}

\def\gsim{\raise0.3ex\hbox{$\;>$\kern-0.75em\raise-1.1ex\hbox{$\sim\;$}}}
\def\lsim{\raise0.3ex\hbox{$\;<$\kern-0.75em\raise-1.1ex\hbox{$\sim\;$}}}

\def\beqn#1{\begin{equation}\label{#1}}
\def\eeqn{\end{equation}}

\def\beqa#1{\begin{eqnarray}\label{#1}}
\def\eeqa{\end{eqnarray}}

\newcommand {\ignore}[1]{}

\def\321{$\mathrm{SU(3) \otimes SU(2) \otimes U(1)}$ }
\newcommand{\vtext}[1]{\begin{sideways}\small{#1}\end{sideways}}

\def\black{\color{black}{}}

\begin{document}

\title{Phenomenology of Dirac neutrino EFTs up to dimension six}

\author{Anirban Biswas}\email{anirban.biswas.sinp@gmail.com}
\affiliation{Department of Physics, School of Sciences and Humanities, SR University, Warangal 506371, India}
\affiliation{Department of Physics, Gaya College (A constituent
unit of Magadh University, Bodh Gaya), Gaya 823001, India}
\author{Eung Jin Chun\orcidlink{0000-0003-4786-313X}}\email{ejchun@kias.re.kr}
\affiliation{Korea Institute for Advanced Study, Seoul 02455, Korea}
\author{Sanjoy Mandal\orcidlink{0000-0003-0171-0752}}\email{smandal@kias.re.kr}
\affiliation{Korea Institute for Advanced Study, Seoul 02455, Korea}
\author{Dibyendu Nanda\orcidlink{0000-0002-7768-7029}}\email{dibyendu.nanda@iopb.res.in}
   \affiliation{Institute of Physics, Sachivalaya Marg, Bhubaneswar 751005, India}
   \affiliation{Homi Bhabha National Institute, Anushakti Nagar, Mumbai 400094, India}
\begin{abstract}
The gauge-singlet right-handed neutrinos would be essential to explain the tiny masses of active neutrinos. We consider the effective field theory of the Standard Model extended with these fields under the assumption that neutrinos are Dirac particles. In this framework, we provide a comprehensive study for the phenomenological consequences of various dimension six interactions employing various high and low energy observables. These include the neutrino mass itself, constraints from electroweak precision test and collider searches for lepton or jet plus missing energy, coherent neutrino-nucleus scattering, beta decays, as well as decays of proton, meson, tau, and top. We also study their astrophysical and cosmological implications for stellar cooling and relativistic degrees of freedom.
\end{abstract}
\maketitle
\section{Introduction}
\label{sec:intro}
Despite the impressive success of the Standard Model~(SM) in describing particle dynamics up to the TeV scale, there are many compelling reasons to believe that it is incomplete. The main experimental indications for new physics are the nonvanishing neutrino masses~\cite{Kajita:2016cak,McDonald:2016ixn,KamLAND:2002uet,K2K:2002icj} and dark matter~\cite{Bertone:2016nfn,deSwart:2017heh,Planck:2018vyg}, which motivate us to construct beyond Standard Model~(BSM) theories that can satisfactorily explain these questions. The BSM theories typically contain new degrees of freedom~(d.o.f), which usually interact with the SM states. Given the null results from various experimental collaborations, these new particles might lie at energies ($\Lambda$) well above the electroweak~(EW) scale. Although the energy of the present day colliders is not sufficient to produce them, the indirect effects of these particles might be detected while analyzing different low-energy observables~\cite{Buchmuller:1985jz}. In view of this, the effective field theory~(EFT) approach \cite{Callan:1969sn,Weinberg:1978kz,Brivio:2017vri,Dobado:1997jx} provides an efficient pathway to parameterize these indirect effects, which can help us uncover the nature of BSM. EFT constructed from SM fields is known as SMEFT~\cite{Buchmuller:1985jz}. This theory adds a tower of effective operators to the renormalizable SM Lagrangian that respects the SM gauge symmetries but not necessarily the global~(accidental) ones. The EFT is valid at
energies below the scale $\Lambda$ where one assumes that the underlying physics is
decoupling~\cite{Appelquist:1974tg,Weinberg:1980wa}, which ensures that all the low-energy observables are suppressed by inverse powers of the cutoff scale $\Lambda$. The EFTs have in common that they predict novel processes that are not present at all in the renormalizable SM.
\par A particularly interesting question in BSM physics is the origin of neutrino masses, which are much smaller than those of all other SM particles, and it is often argued that this smallness could be explained easily if neutrinos are Majorana particles~\cite{Cai:2017jrq}. However, the neutrinos could still be Dirac particles as we have not so far observed neutrinoless double beta decay~($0\nu\beta\beta$)~\cite{KamLAND-Zen:2016pfg,Agostini:2017iyd} or any other lepton number violating~(LNV) processes. In the case of Dirac neutrinos, one needs to add right-handed neutrinos~(RHNs) $N_R$ to the SM particle contents and write down the Yukawa interaction for neutrinos similar to other SM charged fermions. In the case of Dirac/Majorana neutrinos, the SMEFT has to be extended with effective operators containing the RHNs $N_R$. The EFT of this kind is dubbed as $\nu$SMEFT~\cite{delAguila:2008ir}. There are many works which encompass different aspects of $\nu$SMEFT~\cite{delAguila:2008ir,Aparici:2009fh,Bhattacharya:2015vja,Liao:2016qyd,Alcaide:2019pnf,Li:2021tsq,Mitra:2022nri,Mitra:2024ebr,DeVries:2020jbs,Barducci:2022hll,Beltran:2021hpq,Helo:2018bgb,Duarte:2018xst,Zapata:2022qwo,Ardu:2024tzb,Duarte:2020vgj}, but most of these are studied in the context of Majorana neutrinos. Instead, here we focus on the phenomenology of $\nu$SMEFT when neutrinos are Dirac in nature. We will present the low energy phenomenological description of the EFT operators up to dimension six to set constraints coming from the LHC searches for monolepton and monojet plus missing energy, coherent neutrino-nucleus scattering, beta decays, as well as proton, pion, tau, and top decays. 
We also consider the contribution to the number of relativistic species, which depends on the cutoff scale $\Lambda$. It is generally known that RHNs do not contribute to the number of relativistic species, $N_{\rm eff}$, in the absence of any other interactions. But we find that with a low cutoff scale $\Lambda$ in the $\nu$SMEFT, this is no longer true.
\par The rest of this paper is organized as follows. In Sec.~\ref{sec:setup}, we introduce the formalism of the $\nu$SMEFT, listing all possible dimension six operators involving RHNs $N_R$ and discussing the various contributions to the Dirac neutrino masses coming from these EFT operators. In Sec.~\ref{sec:existing_constraints}, we first discuss the existing constraints on the relevant operators coming from various observables. In Sec.~\ref{sec:low_energy_signatures}, we discuss how the low energy signatures of the dimension six operators, such as proton/neutron, meson, tau, and top decays, $\beta$ decays, put constraints on the cutoff scale. In Sec.~\ref{sec:seven}, we discuss the constraints from the COHERENT observation of the CE$\nu$NS process. Further in Sec.~\ref{sec:stellar-colling} and \ref{sec:Neff}, we discuss the effect of these dimension six operators on the stellar cooling and radiation energy density of the early Universe, respectively.  We summarize our findings along with a few concluding remarks in Sec.~\ref{sec:conclusion}.
\section{General setup}
\label{sec:setup}
We provide a brief overview of the $\nu$SMEFT in this section, focusing only on the operators relevant to the current study. We will assume that there are three SM singlet RHNs in addition to the SM particle contents. The most general form of the effective Lagrangian up to dimension six, including these RHNs, is
\begin{align}
\mathcal{L}=\mathcal{L}_{\text{SM}} + i \bar{N}\gamma^\mu\partial_\mu N - \left( Y_\nu \bar{L} \tilde{H} N + \text{h.c.}\right)+\frac{1}{\Lambda^{2}}\sum_{i}^{n} C_i^{(6)}\mathcal{O}_i^{(6)},
\label{eq:Llag}
\end{align}
where $\tilde{H}=i\sigma_2 H^{*}$, $L$ is the SM lepton doublet and $Y_\nu$ is the Dirac-type Yukawa coupling. The $\mathcal{O}_i^{(6)}$ are a set of dimension six operators and are invariant under the SM gauge group $SU(3)_c\otimes SU(2)_L\otimes U(1)_Y$, and $C_i^{(6)}$ are the Wilson coefficient which we assumed to be real. The cutoff scale of the EFT is denoted by $\Lambda$. 
As we are considering Dirac neutrinos, we focus only on $B-L$ invariant dimension six operators~\footnote{Note that the dimension five operators involving RHNs are absent due to the $B-L$ symmetry.}. The basis of dimension six operators involving RHNs $N_R$ were presented in Ref.~\cite{Liao:2016qyd}. 
In Table.~\ref{tab:bosonic} and \ref{tab:fermionic}, we listed possible two-fermion and four-fermion operators, respectively. The four-fermion operators can be categorized into the following types of classes: 
RRRR, LLRR, and LRRL, where L (R) denotes left (right) handed fermions. 
\begin{table}[htb!]
\renewcommand{\arraystretch}{1.5}
\centering
\begin{tabular}{|c | c c|}
\hline
$\psi^2 H X$ & ${\cal O}_{LNB} (+\text{h.c.})=\overline{L}\sigma^{\mu\nu}N\tilde{H}B_{\mu\nu}$  & ${\cal O}_{LNW} (+\text{h.c.})=\overline{L}\sigma^{\mu\nu}N\sigma_I\tilde{H} W^I_{\mu\nu}$ \\
\hline
$\psi^2 H^2 D$  &  ${\cal O}_{HN}=\bar{N}\gamma^\mu N (H^\dagger i \overleftrightarrow{D}_{\mu} H)$  &   ${\cal O}_{HNe} (+\text{h.c.})=\bar{N}\gamma^\mu e_R(\tilde{H}^\dagger i D_\mu H)$ \\
\hline
$\psi^2 H^3$  & \multicolumn{2}{c|}{${\cal O}_{LNH}=\overline{L}\tilde{H} N (H^\dagger H)$}  \\
\hline
\end{tabular}
\caption{List of all possible two-fermion lepton and baryon number conserving operators in the presence of $N_R$ that appear in the dimension six construction. For simplicity, the flavor indices are suppressed. $B_{\mu\nu}$ and $W_{\mu\nu}^I$
represent the weak field strength tensors, and $D_\mu$ is the covariant derivative. Flavor indices are not shown explicitly.}
\label{tab:bosonic}
\end{table}
\begin{table}[htb!]
\renewcommand{\arraystretch}{1.5}
\centering
\begin{tabular}{|c | c c|}
\hline
\multirow{3}{*}{\vtext{RRRR}}&\multicolumn{2}{c|}{$\mathcal{O}_{NN}=(\overline{N}\gamma_\mu N)(\overline{N}\gamma^\mu N)$} \\
&${\cal O}_{eeNN}=(\overline{e_R}\gamma_\mu e_R)(\overline{N}\gamma^\mu N)$&${\cal O}_{uuNN}=(\overline{u_R}\gamma_\mu u_R)(\overline{N}\gamma^\mu N)$\\
&${\cal O}_{ddNN}=(\overline{d_R}\gamma_\mu d_R)(\overline{N}\gamma^\mu N)$&${\cal O}_{duNe} (+\text{h.c.})=(\overline{d_R}\gamma_\mu u_R)(\overline{N}\gamma^\mu e_R)$\\
\hline
LLRR&${\cal O}_{LLNN}=(\overline{L}\gamma_\mu L)(\overline{N}\gamma^\mu N)$&${\cal O}_{QQNN}=(\overline{Q}\gamma_\mu Q)(\overline{N}\gamma^\mu N)$\\
\hline
\multirow{2}{*}{\vtext{LRLR}}&${\cal O}_{LNLe} (+\text{h.c.})=(\overline{L} N)\epsilon (\overline{L}e_R)$&${\cal O}_{LNQd} (+\text{h.c.})=(\overline{L} N)\epsilon (\overline{Q} d_R)$\\
& \multicolumn{2}{c|}{${\cal O}_{LdQN} (+\text{h.c.})=(\overline{L}d_R)\epsilon (\overline{Q} N)$} \\
\hline
LRRL & \multicolumn{2}{c|}{${\cal O}_{QuNL} (+\text{h.c.})=(\overline{Q}u_R)(\overline{N}L)$  }  \\
\hline
\end{tabular}
\caption{List of all possible four-fermion lepton and baryon number conserving operators in the presence of $N_R$ that appear in dimension six construction. Flavor indices are not shown explicitly.}
\label{tab:fermionic}
\end{table}
\par At dimension four, a Dirac neutrino mass term  $m_\nu \overline{\nu_L} N_R$  rises from the operator $Y_\nu\overline{L}\tilde{H}N_R$ after electroweak symmetry breaking   with $m_\nu=Y_\nu v/\sqrt{2}$. Here, the Higgs vacuum expectation value is denoted by $v$. Hence, to be consistent with the present information on the scale of $m_\nu$ requires that $Y_\nu\leq 6\times 10^{-13}$. There will be an additional contribution to the Dirac neutrino mass term $\delta m_\nu$ at dimension six from the operator $\mathcal{O}_{LNH}$, and it can contribute significantly to $\delta m_\nu$ when $\Lambda$ is not significantly bigger than $v$. After electroweak breaking, the operator $\mathcal{O}_{LNH}$ generates the following contribution to neutrino mass,
\begin{align}
\delta m_\nu=C_{LNH}(v)\frac{v^3}{2\sqrt{2} \Lambda^2}.
\label{eq:dm}
\end{align}
Note that under the renormalization group~(RG) running the operators $\mathcal{O}_{LNB,LNW}$ and $\mathcal{O}_{LNH}$ are close in the sense that starting from the Wilson coefficients $C_{LNB,LNW,LNH} (\mu = \Lambda)$ at the scale $\mu=\Lambda$, the RG running leads to mixings between $\mathcal{O}_{LNB,LNW}$ and $\mathcal{O}_{LNH}$ such that $C_{LNH} (\mu = v)$ receives a contribution from $C_{LNB,LNW}(\mu = \Lambda)$~\cite{Bell:2005kz}.   
Assuming the neutrino mass correction $\delta m_\nu\sim 0.1$ eV and $C_{LNH} (\Lambda)\sim 1$, this gives the bound on the cutoff scale as $\Lambda> 2.6\times 10^8$ GeV. We find that even if we take into account the contributions of the operators $\mathcal{O}_{LNB}$ and $\mathcal{O}_{LNW}$, the bound on the cutoff scale does not change much. For example, in the case of $C_{LNB,LNW,LNH} (\Lambda)\sim 1$, the bound is $\Lambda> 2.9\times 10^8$ GeV.
\par The operators such as $\mathcal{O}_{LNB}$ and $\mathcal{O}_{LNW}$ contribute to the following active-sterile transition magnetic moments,
\begin{align}
-\mathcal{L}_{\mu N}=\frac{\mu_{\nu N}}{2} F_{\mu\nu} \overline{\nu_L}\sigma^{\mu\nu} N_R + \text{h.c.}    
\end{align}
In the minimally extended SM~[without the EFT operators in Eq.~\eqref{eq:Llag}], one finds that $\mu_{\nu N}$ is nonvanishing, but unobservably small: $\mu_{\nu N}\approx 3\times 10^{-19}\mu_{B} [m_{\nu}/1\,\text{eV}]$~\cite{Lee:1977tib,Fujikawa:1980yx}. After the electroweak breaking, the magnetic moments result from the combination of $C_{LNB}\mathcal{O}_{LNB}+C_{LNW}\mathcal{O}_{LNW}$~\cite{Brdar:2020quo,Chala:2020pbn},
\begin{align}
\frac{\mu_{\nu N}}{\mu_B}=\frac{2\sqrt{2}m_e v}{e\Lambda^2} \left(c_w C_{LNB}(v) + s_w C_{LNW}(v)\right),
\label{eq:mu}
\end{align}
where $\mu_B=e/2m_e$, $c_w(s_w)=\cos\theta_w~(\sin\theta_w)$ and $\theta_w$ is the Weinberg mixing angle. The cooling of red giant stars plays a prominent role in constraining these magnetic moment operators, which we discuss later. 
Using Eq.~\eqref{eq:dm} and \eqref{eq:mu}, we can write the following relation between $\delta m_\nu$ and $\mu_{\nu N}$,
\begin{align}
\delta m_\nu=\frac{v^2 e}{8m_e}\frac{C_{LNH}(v)}{c_w C_{LNB}(v)+s_w C_{LNW}(v)} \frac{\mu_{\nu N}}{\mu_B}.
\end{align}
It was shown in~\cite{Bell:2005kz} that it is possible to obtain a natural upper bound on $\mu_{\nu N}$ assuming $C_{LNH}(\Lambda)=0$ so that $\delta m_{\nu}$ arises solely from radiative correction involving insertions of $\mathcal{O}_{LNB,LNW}$. With this assumption and setting $\Lambda=1$~TeV, we find the following upper bound,~\footnote{This constraint does not hold true in a number
of general circumstances and is only applicable when $N_R$ is a Weyl field forming a Dirac pair with $\nu_L$~\cite{Bolton:2021pey,Bell:2005kz,deGouvea:2022znk}. Also, this conclusion changes completely if one allows for considerable fine-tuning between $Y_\nu$ and $C_{LNH}$ to make the light neutrino mass of the order of $0.1$~ eV. For a detailed discussion on this, see Ref.~\cite{Chala:2020pbn}.}
\begin{align}
 \frac{\mu_{\nu N}}{\mu_B}\sim 10^{-15} \, \left(\frac{\delta m_{\nu}}{1\,\text{eV}}\right).
 \label{eq:muN}
\end{align}
This bound becomes considerably more stringent as the scale of new physics $\Lambda$ increases from the scale of electroweak symmetry breaking. It can be compared with the present day limits on $\frac{\mu_{\nu N}}{\mu_B}$ derived from solar and reactor neutrinos: $\sim 10^{-10}$~\cite{Beacom:1999wx,Super-Kamiokande:2004wqk,MUNU:2005xnz};  from stellar cooling: $ 3\times 10^{-12}$~\cite{Raffelt:1999gv}; and also from the neutrino-electron scattering: 
$\sim 10^{-10}$~\cite{deGouvea:2022znk}.
\par Lastly we would like to mention that the operators such as $\mathcal{O}_{LNQd}$ and $\mathcal{O}_{QuNL}$ contributes to the Dirac neutrino mass due to the spontaneous breaking of chiral symmetry~(SBCS) via the light quark condensate $\braket{\bar{q}q}=-\Lambda_{\rm QCD}^3$~\cite{Thomas:1992hf,Babic:2019zqu,Davoudiasl:2005ai}. This gives the following contribution to the neutrino mass,
\begin{align}
\delta m_\nu &=-\frac{C_{LNQd}}{\Lambda^2}\braket{\bar{d}d}-\frac{C_{QuNL}}{\Lambda^2} \braket{\bar{u}u}=-\left(C_{LNQd}+C_{QuNL}\right)\frac{\braket{\bar{q}q}}{\Lambda^2}\nonumber \\
& = \left(C_{LNQd}+C_{QuNL}\right)\frac{\Lambda_{\rm QCD}^3}{\Lambda^2},
\label{eq:quark}
\end{align}
where $\braket{\bar{q}q}=\braket{\bar{u}u}=\braket{\bar{d}d}$ and $\Lambda_{\rm QCD}=283$ MeV which we take from a renormalized lattice QCD within the $\overline{\text{MS}}$ scheme at a fixed scale $\mu=2$ GeV~\cite{McNeile:2012xh}. This is a kind of seesaw formula relating the smallness of neutrino mass with the large ratio between the cutoff scale $\Lambda$ and the scale of chiral symmetry breaking $\Lambda_{\rm QCD}$. Taking $C_{LNQd}=C_{QuNL}=1$ and $\delta m_{\nu}<0.1$ eV, Eq.~\eqref{eq:quark} gives a lower bound on the cutoff scale as $\Lambda\geq 21$ TeV.
\par In the following sections, we investigate some phenomenological implications of the assumption that new neutrino interactions are induced by the gauge-invariant dimension-six operators introduced in Tables~\ref{tab:bosonic} and \ref{tab:fermionic}. We will use various collider searches as well as proton/neutron, meson, tau, and top decay measurements to constrain the coefficients or the scale of the $\nu$SMEFT Lagrangian. We also discuss the effect of the dimension six operators on the stellar cooling and cosmological parameter $N_{\rm eff}$.
\section{Constraints from electroweak precision and colliders searches}
\label{sec:existing_constraints}
In this section, we summarize existing limits on some of the dimension six operators. Let us first consider the limits on the bosonic operators listed in Table~\ref{tab:bosonic}. The operators $\mathcal{O}_{LNB,LNW}$ and $\mathcal{O}_{HN}$ trigger the invisible decay channel for the $Z$ boson, whereas the operator $\mathcal{O}_{LNH}$ triggers the invisible decay of Higgs boson~\cite{Barducci:2020icf}. These decay widths are 
\begin{align}
& \Gamma^{\rm NP}(Z\to\text{inv})=\frac{m_Z^3 v^2}{8\pi \Lambda^4}\left(C_{HN}^2+2C_{LNZ}^2\right) ,  \\
&\Gamma^{\rm NP}(h\to\text{inv})=\frac{3m_h v^4}{16\pi\Lambda^4} C_{LNH}^2,
\end{align}
where $C_{LNZ}=c_w C_{LNW}-s_w C_{LNB}$, and flavor indices are suppressed for clarity. LEP experiments place a strong bound on the new physics contributions to the $Z$-boson invisible decay width: $\Gamma^{\rm NP}(Z\to\text{inv})<2$ MeV at $95\%$ C.L~\cite{ALEPH:2005ab}. This gives a constraint on the cutoff scale $\Lambda>1$ TeV, normalizing the relevant Wilson coefficient to one. From the experimental upper bound on the Higgs invisible branching ratio: $\text{BR}(h\to\text{inv})\leq 10.7\%$~\cite{ATLAS:2023tkt}, one obtains $\Lambda>2.8$ TeV with $C_{LNH}= 1$. The operator $\mathcal{O}_{HNe}$ can be constrained by measurements of the $W$ boson width,
\begin{align}
\Delta\Gamma(W\to\ell \nu)=\frac{m_W^3 v^2}{48\pi\Lambda^4} C_{HNe}^2,    
\end{align}
which is bounded as $\Delta\Gamma(W\to\ell \nu)\leq 1.9 \times 10^{-3} \Gamma_W$~\cite{ParticleDataGroup:2024cfk} where $\Gamma_W=2.085$~GeV. 
Thus we get a rather weak bound on the cutoff scale: $\Lambda>0.58$~TeV.

\begin{figure}[!htbp]
\centering
\includegraphics[width=0.28\linewidth]{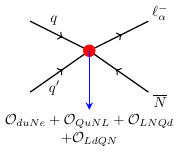}
\includegraphics[width=0.3\linewidth]{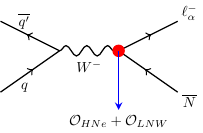}
\includegraphics[width=0.3\linewidth]{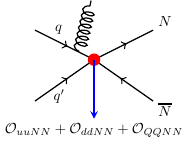}
\includegraphics[width=0.23\linewidth]{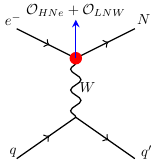}
\includegraphics[width=0.3\linewidth]{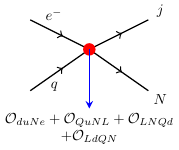}
\caption{The Feynman diagrams and the relevant operators for the processes such as $pp\to\ell^\pm +\slashed{E}_T$~(upper left and middle panel), $pp\to j+\slashed{E}_T$~(upper right panel), and $e^-p\to j+\slashed{E}_T$~(bottom left and right panel). }
\label{fig:Feyn-collider}
\end{figure}
\par The four-fermion operators listed in Table~\ref{tab:fermionic} can have observable consequences for searches at various colliders such as $pp$ and $ep$. In Figure~\ref{fig:Feyn-collider}, we show the Feynman diagrams for possible collider signatures such as $pp\to\ell^\pm +\slashed{E}_T$ (upper left and middle panel), $pp\to j+\slashed{E}_T$ (upper right panel) and $e^-p\to j+\slashed{E}_T$ (bottom left and right panel) with the corresponding operators that contributes to these processes. Although for some of the processes there will be additional contributions from the bosonic operators~($\mathcal{O}_{LNW,HNe}$), we neglect their contributions here for the sake of simplicity. As there is no propagator suppression, the cross section of the signal increases with energy in contrast to the SM background, making the imprint of the four-fermion operators in searches for $\ell+\slashed{E}_T$ more visible in the tail of the lepton's transverse mass distribution. This signal has already been searched for at the LHC \cite{ATLAS:2017jbq}. The detailed methodology of setting bounds on the various operators for the processes $pp\to\ell^\pm +\slashed{E}_T$ is discussed in Ref.~\cite{Alcaide:2019pnf}, where the ATLAS study of Ref.~\cite{ATLAS:2017jbq}, based on 36~$\text{fb}^{-1}$ of data collected at $\sqrt{s}=$13 TeV was considered. The analysis uses events with a high transverse momentum lepton and significant missing transverse momentum. We also considered the final state $\tau+\slashed{E}_T$, which can constrain the same operators with taus instead of light leptons. For this we use the CMS analysis of Ref.~\cite{CMS:2018fza}, based on $35.9\text{ fb}^{-1}$ of data collected at $\sqrt{s}=13$ TeV.

\begin{figure}[!ht]
\centering
\includegraphics[width=0.6\linewidth]{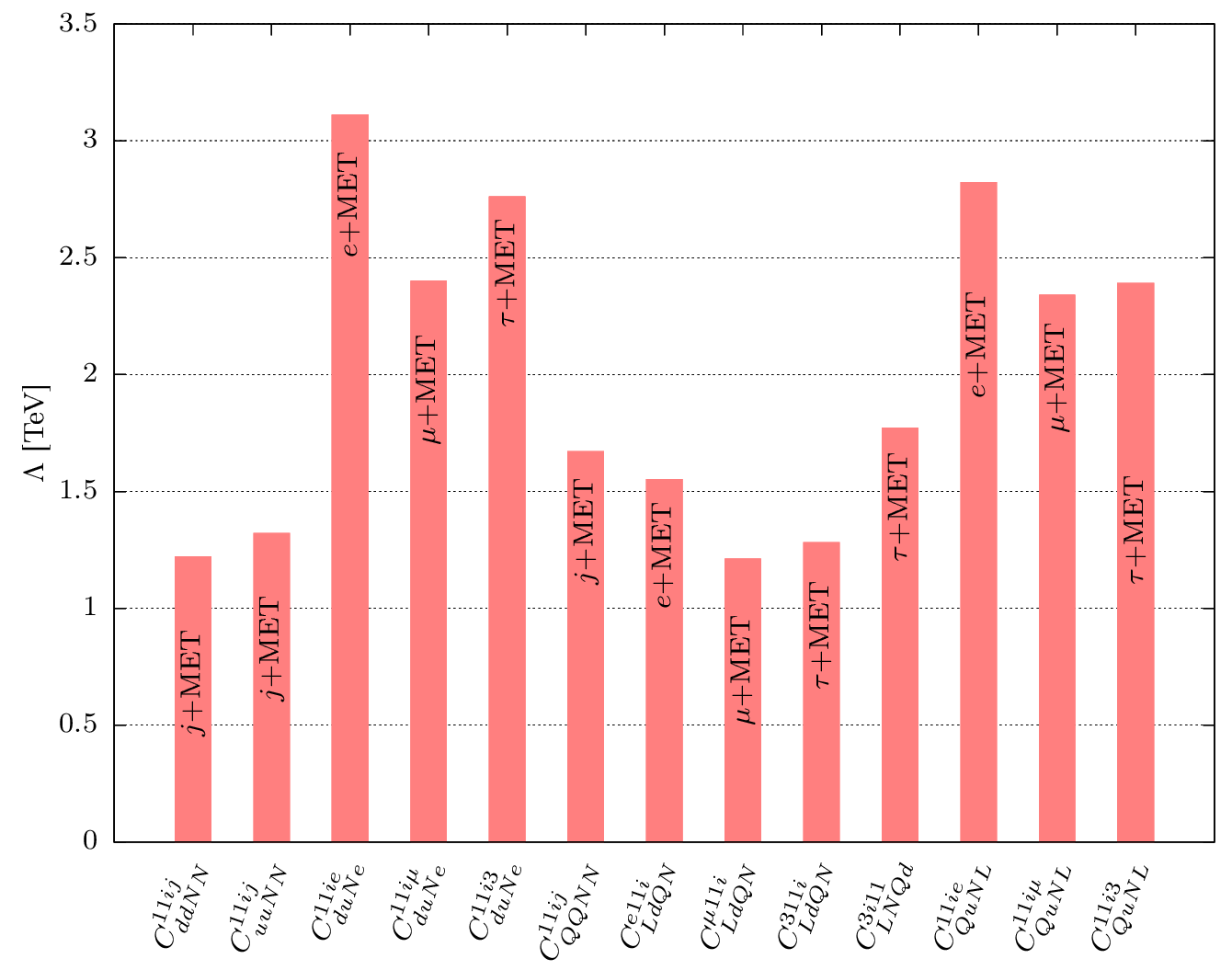}
\caption{Constraints on $\nu$SMEFT cutoff scale $\Lambda$ in unit of TeV, derived from $pp\to\ell+\slashed{E}_T$ and $pp\to j+\slashed{E}_T$. Here, we assume that the Wilson coefficients are of the order of one. The constraints become considerably weaker when the operators are evaluated with second generation quarks rather than first family quarks.}
\label{fig:cons-collider}
\end{figure}

\par In addition to the $\ell+\slashed{E}_T$ final state at $pp$ collider, four fermion operators containing two light quarks and two RHNs ($\mathcal{O}_{uuNN,ddNN,QQNN}$) can lead to monojet signatures at $pp$ collider if for example, a gluon is emitted from one of the initial quarks as shown in the upper right panel of Figure~\ref{fig:Feyn-collider}. 
Following Ref.~\cite{Alcaide:2019pnf}, here we recast the CMS analysis of Ref.~\cite{CMS:2017zts}, based on $35.9\text{ fb}^{-1}$ of data collected at $\sqrt{s}=13$~TeV with the following cuts: 
$\slashed{E}_T>$ 250 GeV, at least one hard jet with $p_T > 250$~GeV and no isolated leptons. 
Note that the $j+\slashed{E}_T$ final state can also arise from the process $pp\to\nu N$ via the operators $O_{QuNL,LdQN}$ and $\mathcal{O}_{LNQd}$. They are constrained also by other observables such as $pp\to\ell+\slashed{E}_T$ and $M_{1}^{+}\to\ell^+ N$ as analyzed in Sec.~\ref{subsec:meson-decay}). Thus, we conservatively neglect them in the analysis of $j+\slashed{E}_T$ final state.
\par The resulting bounds on the cutoff scale coming from the analysis of $j+\slashed{E}_T$ and $\ell+\slashed{E}_T$ are shown in Figure~\ref{fig:cons-collider}. 
In deriving these bounds, we assume only one operator is present at a time, with the corresponding Wilson coefficient normalized to one. Thus, the procedure of setting bounds reduces to one dimension. 
Since the analysis of Ref.~\cite{ATLAS:2017jbq} is more sensitive to electrons than to muons, operators involving electrons are more restricted than those involving muons. Note that although it is possible to constrain the operator $C_{LNQd}^{e/\mu i 11}$ from $e/\mu+\slashed{E}_T$, we find that these are more strongly constrained from pion decay $\pi^{+}\to\ell^+ N$ which will be discussed later. 
\par The same set of operators which induce the process $pp\to\ell+\slashed{E}_T$ also generate the following process $ep\to j+\slashed{E}_T$ and hence can also be constrained from the latter process. But we find the projected bound on the cutoff scale coming from the process $ep\to j+\slashed{E}_T$  is weaker. Although we have not considered this here, the operators such as $\mathcal{O}_{eeNN}$ and $\mathcal{O}_{LLNN}$ can be constrained via the monophoton channel at lepton colliders. 
\section{Low energy signatures of the dimension six operators}
\label{sec:low_energy_signatures}
\subsection{Proton and neutron decay}
In this section, we discuss constraints on various dimension six baryon number violating (BNV) operators that lead to proton and neutron decays. Proton decay is a $|\Delta B|=1$ BNV process that has been predicted by grand unified theories
(GUT) \cite{Pati:1973uk,Georgi:1974sy,Fritzsch:1974nn}. Although BNV has not been observed so far, it is an indispensable ingredient for successful baryogenesis \cite{Sakharov:1967dj}. In SMEFT at dimension six, there appear BNV operators which are invariant under $B-L$ but violate $B+L$ by two units \cite{Weinberg:1979sa,Wilczek:1979hc,Abbott:1980zj,Merlo:2016prs}:
\begin{align}
 \mathcal{O}_{1} =& \left(\overline{{d_{\alpha R}}^c} u_{\beta R}\right) \left(\overline{{Q_{i\gamma L}}^c} L_{jL}\right) \epsilon_{\alpha\beta\gamma} \epsilon_{ij},
 \label{eq:O1}
 \\
 \mathcal{O}_{2} = & \left(\overline{{Q_{i\alpha L}}^c} Q_{j\beta L}\right) \left(\overline{{u_{\gamma R}}^c} e_R\right)\epsilon_{\alpha\beta\gamma} \epsilon_{ij},
 \label{eq:O2}
 \\
 \mathcal{O}_{3} =&\left(\overline{{Q_{i\alpha L}}^c} Q_{j\beta L}\right) \left(\overline{{Q_{k\gamma L}}^c} L_{\ell L}\right) \epsilon_{\alpha\beta\gamma} \epsilon_{ij} \epsilon_{k\ell},
 \label{eq:O3}
 \\
 \mathcal{O}_{4} =&\left(\overline{{d_{\alpha R}}^c} u_{\beta R}\right) \left(\overline{{u_{\gamma R}}^c} e_{R}\right)\epsilon_{\alpha\beta\gamma}.
 \label{eq:O4}
\end{align}
With the addition of singlet RHN, the following two new operators can be written down~\cite{delAguila:2008ir,Alonso:2014zka,Liao:2016qyd,Merlo:2016prs,Helo:2018bgb}:
\begin{align}
&\mathcal{O}_{N1}=\left(\overline{{Q_{i\alpha L}}^c} Q_{j\beta L}\right) \left(\overline{{d_{\gamma R}}^c} N_{R}\right) \epsilon_{\alpha\beta\gamma}\epsilon_{ij},  \\
& \mathcal{O}_{N2} = \left(\overline{{u_{\alpha R}}^c} d_{\beta R}\right) \left(\overline{{d_{\gamma R}}^c} N_{R}\right)\epsilon_{\alpha\beta\gamma}.
\end{align}
Proton/neutron decay modes differ depending on the operators under consideration. The details of the proton/neutron decay calculations are summarized in Appendix~\ref{app:proton_decay}.
\begin{table}[ht]
\begin{center}
\begin{tabular}{ ||c | c | c |c || }
\hline 
\hspace{0.5cm} Process & \hspace{0.05cm} $\tau$~($10^{33}$ years) \hspace{0.05cm} & \hspace{0.05cm} operators \hspace{0.05cm} & $\Lambda_{\text{min}}$~[$10^{15}$ GeV] \\ 
\hline \hline
$p\to\pi^0 e^+$~$(\pi^0\mu^+)$ &  24~(16)~\cite{Super-Kamiokande:2020wjk} &  $C_{1,2,3,4}^{111\ell}$ & 3.88~(3.52) \\
$n\to \pi^- e^+$~$(\pi^-\mu^+)$ &  5.3~(3.5)~\cite{Super-Kamiokande:2017gev}  &           &   2.66~(2.41)\\
\hline
$p\to\eta^0 e^+$~$(\eta^0\mu^+)$ &  10~(4.7)~\cite{Super-Kamiokande:2017gev} &  $C_{1,2,3,4}^{111\ell}$ & 2.56~(2.12)\\
$n\to\eta^0\nu$             &   0.158~\cite{McGrew:1999nd}          &        $C_{1,3}^{111\ell}$,  $C_{N1,N2}^{111i}$       &  0.9 \\
\hline
$p\to K^+\nu$                   &  6.61~\cite{Mine:2016mxy}                 & $C_{1,3,N1}^{211i}, C_{1,3,N1,N2}^{112i}, C_{3,N1,N2}^{121i}$ &   3.46 \\
$n\to K^0\nu$             &   0.13~\cite{Super-Kamiokande:2013rwg}          &                 & 1.3 \\
\hline
$p\to K^0 e^+$~$(K^0\mu^+)$    &    1~(1.6)~\cite{Super-Kamiokande:2005lev,Super-Kamiokande:2012zik}    &  $C_{1,2,3,4}^{211\ell}, C_{2,3}^{121\ell}$ & 1.1~(1.23) \\
\hline
$p\to\pi^+\nu$               &  0.39~\cite{Super-Kamiokande:2013rwg}          &  $C_{1,3,N1,N2}^{111i}$       &        1.64   \\
$n\to\pi^0\nu$             &   1.1~\cite{Super-Kamiokande:2013rwg}          &                  & 2.14\\
\hline                                                                                                                                                                
\end{tabular}
\end{center}
\caption{Allowed two-body decays of nucleons with an upper limit on the lifetime $\tau$ of $90\%$ Confidence level. The third column shows the relevant operators corresponding to each process, and the fourth column shows the corresponding bound on the cutoff scale $\Lambda$, assuming the Wilson coefficients are of order one.}
\label{tab:proton-lifetime}
\end{table} 
\begin{figure}[!ht]
\centering
\includegraphics[width=0.6\linewidth]{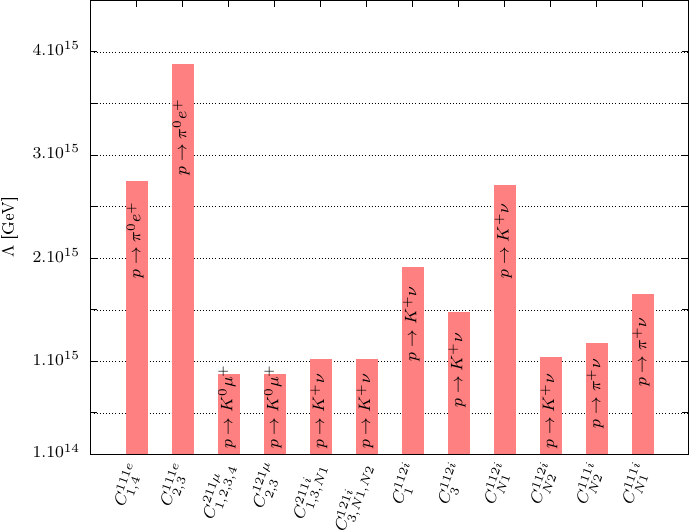}
\caption{Constraints on $\Delta B=1$ $\nu$SMEFT cutoff scale $\Lambda$ in units of GeV, derived from the upper bound on $N\to P\bar{\ell}$ nucleon decay life time, see the second column of Table.~\ref{tab:proton-lifetime}. Here we assume that the Wilson coefficients are of the order of one. }
\label{fig:cons-proton}
\end{figure}
Note that with the above BNV operators, baryon can only decay into an antilepton and a meson, respecting the $B-L$ symmetry. In Table~\ref{tab:proton-lifetime},  we summarize the operators and the associated decay modes along with the present bounds and future sensitivity. The lower indices of the Wilson coefficients $C$ represent the operators $\mathcal{O}_{1,2,3,4,N1,N2}$ introduced in the above equations, and the upper indices denote different flavors for given operators. 
We find that assuming the involving Wilson coefficient operators to be of the order one, the upper bound on the proton lifetime translates into a lower bound on the cutoff scale $\Lambda$ of about $10^{15}$ GeV, see the fourth column of Table~\ref{tab:proton-lifetime}. Note that the bound on the cutoff scale mildly depends on the specific flavor structure, which we listed in Figure~\ref{fig:cons-proton}. The tightest constraints come from the decay mode $p\to \pi^0 e^+$ as this mode is more strongly constrained compared to other modes. 
\par The presence of the additional BNV operators involving RHNs $N_R$ introduces new decay channels for proton and neutron. Hence, observation/nonobservation of particular decay modes might hint at the existence of RHNs~\cite{Helo:2018bgb}. The operators $\mathcal{O}_{1, 3}$ and $\mathcal{O}_{N1,N2}$ can produce the final state, which has a charged pion~($\pi^+$) and missing energy. From the argument of isospin symmetry, $\mathcal{O}_{1,3}$ also generates the process with the left-handed charged lepton, which is $p\to\pi^0\ell^+$. On the other hand, the operators $\mathcal{O}_{N1,N2}$ cannot generate the decay mode $p\to\pi^0\ell^+$ since these operators do not have charged lepton counterparts. Hence, observation of proton decay with a final state of $p\to\pi^+ + \slashed{E}$ and the simultaneous absence of the $\pi^0\ell^+$ mode hints the existence of RHNs $N_R$ which act as a Dirac partner of the ordinary neutrino. The same logic also holds for neutron decay. More specifically, the operators $\mathcal{O}_{N1,N2}$ can generate neutron decay $n\to\pi^0+\slashed{E}$ but not the $n\to\pi^- e^+$. Hence, observation of $n\to\pi^++\slashed{E}$ can be interpreted as the existence of RHNs~\footnote{This argument is only valid if $SU(2)_L$ invariance is preserved, which we assumed in the construction of all nonrenormalizable BNV operators.}. This also suggests that the proton lifetime might be related to the nature of ordinary neutrinos. For a detailed discussion on the relation between the proton decay mode $p\to\pi^++\slashed{E}$ and Dirac neutrino mass models with the full decomposition of the proton decay operators $\mathcal{O}_{N1,N2}$, see Refs.~\cite{Helo:2019yqp,Helo:2018bgb}.
\subsection{Meson decay: $ M_1^\pm\to \ell^\pm N$}
\label{subsec:meson-decay}
In addition to contributing to the $pp\to\ell N$ process as described in Sec.~\ref{sec:existing_constraints}, the operators $\mathcal{O}_{duNe}, \mathcal{O}_{LNQd}$, $\mathcal{O}_{LdQN}$ and $\mathcal{O}_{QuNL}$ that produce four-point interactions of two quarks, a lepton, and a RHN, also contributes to the meson decay $M_1^\pm\to\ell^\pm N$ \footnote{We neglect the contribution coming from the operator $\mathcal{O}_{LdQN}$ as it is rather difficult to estimate the meson form factor.}. Apart from these four-fermion operators, there will be additional contributions from the two-fermion operators such as $\mathcal{O}_{HNe}$ and $\mathcal{O}_{LNW}$. 

\begin{figure}[!htbp]
\centering
\includegraphics[width=0.4\linewidth]{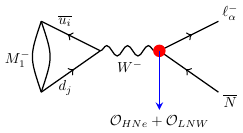}~~~~
\includegraphics[width=0.35\linewidth]{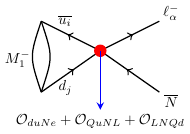}
\caption{The Feynman diagrams and the relevant operators for the meson decay $M_{1}^-\to\ell_\alpha^-\overline{N}$. The left and right panel stands for contributions coming from two-fermion and four-fermion operators.}
\label{fig:Feyn-meson}
\end{figure}

Figure~\ref{fig:Feyn-meson} shows the Feynman diagrams corresponding to the relevant operators from which one can calculate the decay amplitude as
\begin{align}
\mathcal{M}\left(M_1^-(\overline{u_i}d_j)\to\ell_\alpha^- \overline{N}\right)&=\frac{V_{u_{i}d_{j}}^{\rm CKM}}{2\Lambda^2} f_{M_1}\Bigg\{\left(C_{duNe}^{jiN\alpha}-C_{HNe}^{N\alpha}\right)\overline{u}(\ell_\alpha)\slashed{p}P_R v(\overline{N})+\frac{2C_{LNW}^{\alpha N}}{M_W}\overline{u}(\ell_\alpha)\sigma^{\alpha\beta}P_R v(\overline{N}) p_\alpha p_\beta \nonumber \\
& + \frac{m_{M_1}^2}{(m_{u_i}+m_{d_j})}\left(C_{QuNL}^{jiN\alpha}-C_{LNQd}^{\alpha Nij}\right)\overline{u}(\ell_\alpha)P_R v(\overline{N})\Bigg\},
\end{align}
where $p=p_{\ell_\alpha}+p_N$. A similar expression holds for $M_{1}^+\to\ell^+ N$.
Following Ref.~\cite{Carpentier:2010ue}, we use
 $\bra{0}V^\mu\ket{M_1^\pm}=f_{M_1}p^\mu$ and $\bra{0} S\ket{M_1^\pm}=f_{M_1}\frac{M_1^2}{(m_{u_i}+m_{d_j})}$. With this, the corresponding decay width reads
\small
\begin{align}
\Gamma\left(M_1^-(\overline{u_i}d_j)\to\ell_\alpha^- \overline{N}\right)=\frac{|V_{u_{i}d_{j}}^{\rm CKM}|^2}{32\pi \Lambda^4}f_{M_1}^2 m_{M_1}\left(1-\frac{m_{\ell_\alpha}^2}{m_{M_1}^2}\right)^2\Big[m_{\ell_\alpha}(C_{duNe}^{jiN\alpha}-C_{HNe}^{N\alpha})+\frac{m_{M_1}^2}{(m_{u_i}+m_{d_j})}(C_{QuNL}^{jiN\alpha}-C_{LNQd}^{\alpha Nij})\Big]^2.
\end{align}
\normalsize
\begin{table}[ht]
\begin{tabular}{ ||c | c | c|c  || }
\hline 
\hspace{0.5cm} Meson decay & \hspace{0.05cm} Decay width [GeV] \hspace{0.05cm} & \hspace{0.05cm} Relevant Coefficients \hspace{0.05cm}  & $\Lambda_{\rm min}$ [TeV]\\ 
\hline \hline
$\pi^+\to\mu^+\nu_\mu $ & $(2.5281\pm 0.0005)\times 10^{-17}$  &   $C_{duNe}^{111\mu}$, $C_{HNe}^{i\mu}$, $C_{QuNL}^{11i\mu}$, $C_{LNQd}^{\mu i 11}$ & 2.45\\
$\pi^+\to e^+\nu_e $ & $(3.110\pm 0.010)\times 10^{-21}$  &   $C_{duNe}^{111e}$, $C_{HNe}^{ie}$, $C_{QuNL}^{11ie}$, $C_{LNQd}^{ei 11}$ & 1.23 \\
\hline
$K^+\to\mu^+\nu_\mu $ & $(3.379\pm 0.008)\times 10^{-17}$  &   $C_{duNe}^{21i\mu}$, $C_{HNe}^{i\mu}$, $C_{QuNL}^{21i\mu}$, $C_{LNQd}^{\mu i 12}$ & 1.32\\
$K^+\to e^+\nu_e $ & $(8.41\pm 0.04)\times 10^{-22}$  &     $C_{duNe}^{21ie}$, $C_{HNe}^{ie}$, $C_{QuNL}^{21ie}$, $C_{LNQd}^{e i 12}$ & 1.12\\
\hline      
$ D_s^+\to\mu^+\nu_\mu $ & $(7.09\pm 0.20)\times 10^{-15}$  &   $C_{duNe}^{22i\mu}$, $C_{HNe}^{i\mu}$, $C_{QuNL}^{22i\mu}$, $C_{LNQd}^{\mu i 22}$  & 0.73\\
$ D_s^+\to\tau^+\nu_\tau $ & $(6.95\pm 0.15)\times 10^{-14}$ &   $C_{duNe}^{22i\tau}$, $C_{HNe}^{i\tau}$, $C_{QuNL}^{22i\tau}$, $C_{LNQd}^{\tau i 22}$  & 0.77 \\
\hline 
$ B^+\to\tau^+\nu_\tau $ & $(4.4\pm 1.0)\times 10^{-17}$  &   $C_{duNe}^{31i\tau}$, $C_{HNe}^{i\tau}$, $C_{QuNL}^{31i\tau}$, $C_{LNQd}^{\tau i 13}$  &  0.44\\
\hline 
$ D^+\to\mu^+\nu_\mu $ & $(2.38\pm 0.11)\times 10^{-16}$  &  $C_{duNe}^{12i\mu}$, $C_{HNe}^{i\mu}$, $C_{QuNL}^{12i\mu}$, $C_{LNQd}^{\mu i 21}$   &  0.63 \\
$ D^+\to\tau^+\nu_\tau $ & $(7.6\pm 1.7)\times 10^{-16}$ &   $C_{duNe}^{12i\tau}$, $C_{HNe}^{i\tau}$, $C_{QuNL}^{12i\tau}$, $C_{LNQd}^{\tau i 21}$ &   0.72 \\
\hline                                            
\end{tabular}
\caption{The measured decay width with corresponding uncertainties for $M_1^+\to\ell^+ + \slashed{E}$ decay mode. In the third column, we listed the relevant operators, and the fourth column shows the corresponding lower bound on the cutoff scale $\Lambda$, assuming the involved Wilson coefficients to be equal and order one.}
\label{tab:decay-width-meson}
\end{table} 
The experimental values of the various meson decay widths to $\ell+\slashed{E}$ with the corresponding uncertainty are listed in Table~\ref{tab:decay-width-meson}, which we have taken from Ref.~\cite{ParticleDataGroup:2024cfk}. We set the bound on the Wilson coefficients or the cutoff scale entering this equation by requiring that the corresponding contribution is smaller than twice the experimental error. The resulting bounds on the cutoff scale coming from the various meson decays to $\ell^++\slashed{E}$ are listed in the fourth column of Table~\ref{tab:decay-width-meson}.
 \par There can also be constraints on the Wilson coefficients from meson invisible decays~\cite{ParticleDataGroup:2018ovx}. In the case of a pseudoscalar meson $P$, the vacuum-to-meson transition matrix elements of the scalar, vector, and tensor quark currents vanish identically. Hence only operators such as $\mathcal{O}_{LNQd}$, $\mathcal{O}_{LdQN}$ and $\mathcal{O}_{QuNL}$ contributes. In the case of vector meson $V$, vector~($\mathcal{O}_{uuNN},\mathcal{O}_{ddNN}$ and $\mathcal{O}_{QQNN}$), tensor~($\mathcal{O}_{LdQN}$) and dipole~($\mathcal{O}_{LNB/LNW}$) operators can contributes. For a detailed discussion on invisible meson decays in the context of effective theory see Ref.~\cite{Li:2020lba}. We found that the invisible meson decays typically gives bound on the cutoff scale  as $\Lambda\sim \mathcal{O}(100)$ GeV which is weak compare to the bound coming from leptonic decay of mesons.
\par There could be complementary constraints on effective operators from low energy nuclear-level processes like $\beta$ decays. The operators such as $\mathcal{O}_{QuNL}$, $\mathcal{O}_{LNQd}$ and $\mathcal{O}_{LdQN}$ contributes to the $\beta$ decay. Following Ref.~\cite{Bischer:2019ttk}, we found that the tightest bound on the cutoff scale is $\Lambda\sim 9$ TeV assuming only one of the operators is present at a time.
\black 
\subsection{Tau decays: $\tau\to \ell + \slashed{E}_T$ and $\tau\to M_1+\slashed{E}_T$}
\begin{figure}[!htbp]
\centering
\includegraphics[width=0.3\linewidth]{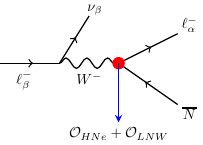}
\includegraphics[width=0.3\linewidth]{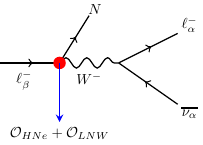}
\includegraphics[width=0.23\linewidth]{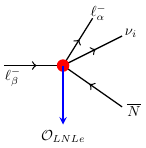}
\includegraphics[width=0.23\linewidth]{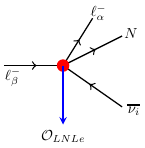}
\includegraphics[width=0.23\linewidth]{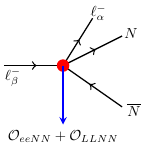}
\caption{The Feynman diagrams and the relevant operators for the decay $\ell_\beta^-\to\ell_\alpha^{-}+\slashed{E}$. }
\label{fig:Feyn-lepton}
\end{figure}

The dimension six operators can also modify the leptonic decay width as shown in Figure~\ref{fig:Feyn-lepton}. The relevant operators involved are the following: $\mathcal{O}_{HNe/LNW}$, $\mathcal{O}_{LNLe}$ and $\mathcal{O}_{LLNN,eeNN}$. The details of the decay width calculations are given in Appendix.~\ref{app:tau-decays}. The decay mode $\mu\to e+\text{inv}$ is precisely measured so that room for new physics is very much restricted. As a result, we consider here tau decays to electron and muon to constrain the involving operators.
\begin{table}[ht]
\begin{tabular}{ ||c | c|c  || }
\hline 
\hspace{0.5cm} Tau decays & \hspace{0.05cm} Relevant Coefficients \hspace{0.05cm}  & $\Lambda_{\rm min}$ [TeV]\\ 
\hline \hline
$\tau^-\to\ell^-+\slashed{E}$   & $C_{HNe}^{i\ell/i\tau}$, $C_{LNW}^{\ell i/\tau i}$, $C_{LNLe}^{\tau i\ell(\tau/\ell)}, C_{LNLe}^{\ell i\tau(\tau/\ell)}, C_{LNLe}^{ij(\ell\tau/\tau\ell)},C_{LNLe}^{lji\tau}, C_{LNLe}^{\tau ji\ell}$, $C_{eeNN/LLNN}^{\ell\tau i j}$  &   1.5~(1.49) \\
\hline                                            
\end{tabular}
\caption{The constraint on the cutoff scale $\Lambda$ from measured values of the tau decay widths into electrons and muons. The second column shows the involving operators, and the third column shows the corresponding lower bound on the cutoff scale, assuming the involved Wilson coefficients to be equal and of order one.}
\label{tab:tau-leptonic-decay}
\end{table} 
The measured values of the tau decay widths into electrons and muons are $\Gamma(\tau\to\mu+\slashed{E})=(3.943\pm 0.011)\times 10^{-13}$ GeV and $\Gamma(\tau \to e+\slashed{E})=(4.04\pm 0.011)\times 10^{-13}$ GeV~\cite{ParticleDataGroup:2024cfk}, respectively.
\begin{table}[ht]
\begin{tabular}{ ||c | c | c|c  || }
\hline 
\hspace{0.5cm} Tau decays &  Decay width [GeV] & \hspace{0.05cm} Relevant Coefficients \hspace{0.05cm}  & $\Lambda_{\rm min}$ [TeV]\\ 
\hline \hline
$\tau \to \pi+\text{inv}$ &  $(2.453\pm 0.012)\times 10^{-13}$  & $C_{duNe}^{11i\tau}, C_{QuNL}^{11i\tau}, C_{LNQd}^{\tau i11}$ &  1.04 \\
\hline
$\tau \to K+\text{inv}$   & $(1.578\pm 0.023)\times 10^{-14}$   & $C_{duNe}^{21i\tau}, C_{QuNL}^{21i\tau}, C_{LNQd}^{\tau i12}$ &   1.65 \\
\hline                                                  
\end{tabular}
\caption{Constraint on the cutoff scale coming from the semileptonic tau decay mode assuming theoretical decay width does not exceed twice the experimental error, and all the involved Wilson coefficients are order one.}
\label{tab:tau-semileptonic-decay}
\end{table} 
In addition to the leptonic decay, we find that one can also have semileptonic decay mode such as $\tau^-\to M_1^- N$~($M_1^-=\pi^-, K^-$) due to the operators $\mathcal{O}_{duNe},\mathcal{O}_{QuNL}$ and $\mathcal{O}_{LNQd}$. The expression for the corresponding decay width is given as
\begin{align}
\Gamma(\tau^-\to P^-(\bar{u_i} d_j) N)=\frac{f_P^2}{32\pi\Lambda^4} m_\tau \left(1-\frac{m_P^2}{m_\tau^2}\right)^2 \Bigg[\left(C_{duNe}^{jiN\tau}\right)^2 m_\tau^2 + \left(C_{QuNL}^{jiN\tau}-C_{LNQd}^{\tau Nij}\right)^2\left(\frac{m_P^2}{m_{u_i}+m_{d_j}}\right)^2\Bigg].
\end{align}
We consider the decay modes $\tau^-\to \pi^-\nu_\tau$ and $\tau^-\to K^-\nu_\tau$ with their measured values as $(2.453\pm 0.012)\times 10^{-13}$ and $(1.578\pm 0.023)\times 10^{-14}$~\cite{ParticleDataGroup:2024cfk}, respectively. For both the tau leptonic and semileptonic decays, we bound the relevant operators, as in the pion case, by requiring that the corresponding theoretical decay width not exceed twice the experimental error, which are shown in Table~\ref{tab:tau-leptonic-decay} and Table~\ref{tab:tau-semileptonic-decay}, respectively.
\subsection{Top decay: $ t\to b\ell N$}
\begin{figure}[!htbp]
\centering
\includegraphics[width=0.38\linewidth]{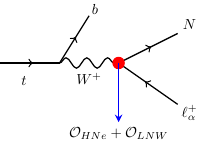}
\includegraphics[width=0.45\linewidth]{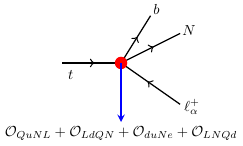}
\caption{The Feynman diagrams and the relevant operators for the top decay mode $t\to b\ell_\alpha^+ N$.}
\label{fig:Feyn-top}
\end{figure}

Dimension six operators can also be probed in top decays. These operator induces new top decays such as  $j+\slashed{E}_T$, $t\to b\ell+\slashed{E}_T$ and $b\tau+\slashed{E}_T$. The decay mode $j+\slashed{E}_T$ is induced by the following dimension six operators: $\mathcal{O}_{uuNN}$, $\mathcal{O}_{QQNN}$ and $\mathcal{O}_{QuNL}$. The corresponding decay widths read
\begin{align}
&\Gamma(t\to u\overline{N}\nu_i)=\frac{m_t^5}{6144\pi^3\Lambda^4}    (C_{QuNL}^{31Ni})^2,\\
&\Gamma(t\to u\overline{\nu_i} N)=\frac{m_t^5}{6144\pi^3\Lambda^4}    (C_{QuNL}^{13Ni})^2,\\
& \Gamma(t\to u\overline{N} N)=\frac{m_t^5}{1536\pi^3\Lambda^4}    \Big[(C_{uuNN}^{13NN})^2+(C_{QQNN}^{13NN})^2\Big].
\end{align}
Similar expressions hold, of course, for second generation quarks. Due to the two sources of missing energy and the light jets involved, this mode is difficult to investigate at hadron colliders. For the decay mode $t\to b\ell+\slashed{E}_T$, the corresponding Feynman diagrams involving the relevant operators are shown in Figure~\ref{fig:Feyn-top}. The relevant amplitude and decay widths read as
\begin{align}
&\mathcal{M}(t\to b\ell_\alpha^+ N) =\frac{V_{tb}^{\rm CKM}}{\Lambda^2}\Big[ C_{HNe}^{N\alpha}\, \overline{u}(b)\gamma^\mu P_L u(t)\,\, \overline{u}(N)\gamma_\mu P_R v(\ell_\alpha^+) -2\frac{C_{LNW}^{\alpha N}}{M_W}\, (p-k_1)_{\mu} \overline{u}(b)\gamma_\nu P_L u(t)\,\, \overline{u}(N)\sigma^{\mu\nu} P_L v(\ell_\alpha^+) \nonumber \\
& + C_{QuNL}^{33N\alpha}\, \overline{u}(b)P_R u(t)\,\, \overline{u}(N) P_L v(\ell_\alpha^+) - C_{LdQN}^{\alpha 33N}\, \overline{u}(b) P_L v(\ell_\alpha^+)\,\, \overline{u}(N) P_L u(t) + C_{duNe}^{33N\alpha}\, \overline{u}(b)\gamma^\mu P_R u(t)\,\, \overline{u}(N) \nonumber \\ 
& \gamma_\mu P_R v(\ell_\alpha^+)
 - C_{LNQd}^{\alpha N33}\, \overline{u}(N)P_L v(\ell_\alpha^+)\,\, \overline{u}(b)P_L u(t)\Big],\\
 &\Gamma(t\to b\ell_\alpha^+ N) = \frac{m_t^5 |V_{tb}^{\rm CKM}|^2}{6144\pi^3\Lambda^4} \Big\{(C_{duNe}^{33N\alpha})^2 + (C_{HNe}^{N\alpha})^2 + \frac{1}{4} (C_{LdQN}^{\alpha 33N})^2 +(C_{LNQd}^{\alpha N 33})^2 + (C_{QuNL}^{33N\alpha})^2 + C_{LNQd}^{\alpha N33} C_{LdQN}^{\alpha 33N}  \Big\}.
\end{align}
Similar expression holds for $\overline{t}\to\overline{b}\ell_\alpha^-\overline{N}$. Note that the contribution from the operators $\mathcal{O}_{HNe}$ and $\mathcal{O}_{LNW}$ vanishes in the massless limit of the final states.
The measured values of the top decay widths into electron, muon, and tau modes are $\Gamma(t\to eb+\text{inv})=0.158_{-0.017}^{+0.022}$ GeV, $\Gamma(t\to \mu b+\text{inv})=0.162_{-0.017}^{+0.022}$ GeV and $\Gamma(t\to \tau b+\text{inv})=0.152_{-0.018}^{+0.022}$ GeV, respectively.
\begin{table}[ht]
\begin{tabular}{ ||c | c|c  || }
\hline 
\hspace{0.5cm} Top decays & \hspace{0.05cm} Relevant Coefficients \hspace{0.05cm}  & $\Lambda_{\rm min}$ [TeV]\\ 
\hline \hline
$t\to\ell_{\alpha}b+\text{Inv}$   & $C_{HNe}^{i\alpha}$, $C_{duNe}^{33i\alpha}, C_{LdQN}^{\alpha 33i}, C_{LNQd}^{\alpha i 33}, C_{QuNL}^{33i\alpha}$ &  0.139, 0.139, 0.137 \\
\hline                                                  
\end{tabular}
\caption{Constraint on the cutoff scale coming from the semileptonic top decay mode assuming theoretical decay
width does not exceed twice the experimental error, and all the involved Wilson coefficients are order one.}
\label{tab:decay-width-top}
\end{table} 
The corresponding bound on the cutoff scale is shown in the Table~\ref{tab:decay-width-top}. The bound on the cutoff scale is rather loose, as the current bounds on the top width are not constraining enough. It was pointed in Ref.~\cite{Alcaide:2019pnf} that one can get tighter bound on the cutoff scale following a search strategy based on a new rare top decay at the LHC~\footnote{Note that the final state $t\to b\ell^+ +\slashed{E}_T$ is also there in the SM, but interestingly in this case with dimension six operators, the lepton and the missing energy do not reconstruct a $W$ boson.}. Reference~\cite{Alcaide:2019pnf} uses the $t\bar{t}$ production, where one of the tops decays leptonically through the modified vertex, while the other tops decay identically as in the SM in the hadronic mode. Considering only the muonic channel, this translates to a prospective lower bound on $\Lambda\sim$~330 (460) GeV for the Wilson coefficient $C\sim 1$. These numbers can rise up to $\Lambda\sim$ 1.8 (2.5) TeV if both
electrons and muons, as well as three RHNs, are included.
\section{Neutrino-nucleus scattering}
\label{sec:seven}
In this section, we study the current bounds from the Coherent elastic neutrino-nucleus scattering (CE$\nu$NS) process~($\overset{\scriptscriptstyle(-)}{\nu} \mathcal{N}\to X \mathcal{N}$, where $X\in \{\nu,\bar\nu, N, \bar N\}$) on the corresponding effective operators.
CE$\nu$NS is a neutral-current process in which a low-energy neutrino scatters off an entire nucleus~\cite{Abdullah:2022zue}. Its experimental detection presents technological challenges, as it involves observing nuclear recoils with extremely low energy. Consequently, the process remained undetected for decades until the COHERENT Collaboration~\cite{COHERENT:2017ipa} first observed it using a spallation source producing neutrinos from pion decay at rest.
Additional observations using various targets~\cite{COHERENT:2020iec, COHERENT:2021xmm, COHERENT:2024axu} and a reactor source~\cite{Colaresi:2022obx} have offered important insights into the CE$\nu$NS cross section. The SM predicts the CE$\nu$NS process via $Z$ boson exchange~\cite{Freedman:1973yd}, and the observed results so far align with the SM within $1\sigma$ uncertainty. In addition to active neutrinos produced via the SM weak neutral current, the CE$\nu$NS process can also yield any neutrino flavor, including light RHNs, in the final state. The COHERENT observation therefore offers an opportunity to investigate new physics (NP) related to general neutrino interactions involving light RHNs.
\par The relevant $\nu$SMEFT operators which contributes to the $\overset{\scriptscriptstyle(-)}{\nu} \mathcal{N}\to X \mathcal{N}$  coherent scattering are ~\cite{Li:2020lba,Bischer:2019ttk},
\begin{align}
-\mathcal{L} &=\frac{(\mu_{\nu N})_{\alpha\beta}}{2}\bar{\nu}_\alpha\sigma^{\mu\nu}P_R N_\beta F_{\mu\nu}+\text{h.c.} + \sqrt{2}G_F \Bigg\{\sum_{q=u,d}\Big[ (g_{V}^{qq})_{\alpha\beta}\, (\bar{q}\gamma^\mu q)\, (\bar{\nu}_\alpha\gamma_\mu P_L \nu_\beta)    + (g_{A}^{qq})_{\alpha\beta} (\bar{q}\gamma^\mu\gamma^5 q)\, (\bar{\nu}_\alpha\gamma_\mu P_L \nu_\beta) \nonumber \\
&+ \Big[ (\tilde{g}_{S}^{dd})_{\alpha\beta} (\overline{d_L} d_R)\, (\bar{\nu}_\alpha P_R N_\beta) + (\tilde{g}_{S}^{uu})_{\alpha\beta} (\overline{u_R} u_L)\, (\bar{\nu}_\alpha P_R N_\beta) + (\tilde{g}_{T}^{dd})_{\alpha\beta}(\overline{d}\sigma^{\mu\nu}P_R d)\, (\bar{\nu}_\alpha\sigma_{\mu\nu}P_R N_\beta) + \text{h.c}\Big]\Bigg\}
\end{align}
The tree-level values of the SM couplings are $g_V^{uu}=1/2(1-8/3\sin^2\theta_W)$, $g_V^{dd}=-1/2(1-4/3\sin^2\theta_W)$, $g_A^{uu}=-1/2$, $g_A^{dd}=1/2$ where $\theta_W$ is the weak mixing angle. We use the tilde to mark the coefficients of dimension six operators that involve $N_R$, which are thus new with respect
to SM. These couplings can be written in terms of Wilson coefficients as,
\begin{align}
(\tilde{g}_{S}^{dd})_{\alpha\beta}=\frac{1}{2\sqrt{2} G_F\Lambda^2}\Big(C_{LNQd}^{\alpha\beta 11}-2C_{LdQN}^{\alpha 1 1\beta}\Big),\,\,
(\tilde{g}_{S}^{uu})_{\alpha\beta}=\frac{1}{\sqrt{2} G_F\Lambda^2} C_{QuNL}^{11\alpha\beta}, \,\,(\tilde{g}_{T}^{dd})_{\alpha\beta}=-\frac{1}{8\sqrt{2} G_F\Lambda^2}C_{LdQN}^{\alpha 1 1\beta}.
\end{align}
A nucleon-level effective Lagrangian offers a convenient intermediate step for describing CE$\nu$NS interactions. The relevant neutral current (NC) interactions  are
\begin{align}
-\mathcal{L} &=\frac{(\mu_{\nu N})_{\alpha\beta}}{2}\bar{\nu}_\alpha\sigma^{\mu\nu}P_R N_\beta F_{\mu\nu}+\text{h.c.} + \frac{G_F}{\sqrt{2}} \Bigg\{ \Big[ (g_V^{\nu\mathcal{N}})_{\alpha\beta} (\overline{\mathcal{N}}\gamma^\mu \mathcal{N})\, (\bar{\nu}_\alpha\gamma_\mu P_L\nu_\beta) + (g_A^{\nu\mathcal{N}})_{\alpha\beta} (\overline{\mathcal{N}}\gamma^\mu\gamma^5 \mathcal{N})\, (\bar{\nu}_\alpha\gamma_\mu P_L\nu_\beta) \Big] \nonumber \\
& + \Big[(\tilde{g}_S^{N\mathcal{N}})_{\alpha\beta} (\overline{\mathcal{N}} \mathcal{N})\, (\overline{\nu}_\alpha P_R N_\beta) + (\tilde{g}_T^{N\mathcal{N}})_{\alpha\beta} (\overline{\mathcal{N}} \sigma^{\mu\nu} P_R \mathcal{N})\, (\overline{\nu}_\alpha \sigma_{\mu\nu} P_R N_\beta) +\text{h.c}\Big] \Bigg\},
\end{align}
where the coefficients,
\begin{align}
& g_V^{\,\nu \mathcal{N}} = 2\mathbb{Z}_i (2 g_V^{uu} + g_V^{dd} ) F_p(q^2) + 2\mathbb{N}_i (g_V^{uu} + 2 g_V^{dd}) F_n(q^2), \\
\label{eq:gV}
& \tilde{g}_S^{\, N\mathcal{N}}=2\sum_{q=u,d} \tilde{g}_S^{\, qq}\Big[\mathbb{Z}_i\frac{m_p}{m_q} f_{T_q}^p F_p(q^2)+ \mathbb{N}_i\frac{m_n}{m_q} f_{T_q}^n F_n(q^2)\Big],\\
& \tilde{g}_T^{\, N\mathcal{N}}=2\sum_{q=u,d} \tilde{g}_T^{\, q q} \Big[\mathbb{Z}_i \delta_q^p F_p(q^2) + \mathbb{N}_i \delta_q^n F_n(q^2)\Big],
\label{eq:gT}
\end{align}
parametrize the vector, scalar and tensor contributions. $\mathbb{Z}_i$ and $\mathbb{N}_i$ are the numbers of protons and neutrons in the nucleus, whereas the $F_p/F_n$ are the proton/neutron form factor. The hadronic structure parameters for the case of scalar interactions: $f_u^p=0.0208$, $f_u^n=0.0189$, $f_d^p=0.0411$,  $f_d^n=0.0451$ and tensor interactions: $\delta^p_u=\delta^n_d=0.54$, $\delta^p_d=\delta^n_u=-0.23$ are taken from Ref.~\cite{AristizabalSierra:2019zmy}. We assume that the form factors of the proton and neutron are both given by the Helm form factor, i.e. $F_p(q^2)=F_n(q^2)=F(q^2)$.
The differential cross section for $\overset{\scriptscriptstyle(-)}{\nu} \mathcal{N}\to X \mathcal{N}$  coherent scattering, is at leading order given by~\cite{Lindner:2016wff,Chang:2020jwl,Li:2020lba,DeRomeri:2022twg,DeRomeri:2024iaw}~\footnote{The interference terms are suppressed by $T/E_\nu$  and are thus not included here.}
\begin{align}
\frac{d\sigma}{dT}=\frac{G_F^2 M}{4\pi}\Bigg[ \xi_S^2\frac{T}{T_{\rm max}}+ \xi_V^2 \left(1-\frac{T}{T_{\rm max}}\right)+ \xi_T^2 \left(1-\frac{T}{2T_{\rm max}}\right) + e^2 A_M^2 \left(\frac{1}{MT}-\frac{1}{ME_\nu}\right)\Bigg],
\label{eq:ds}
\end{align}
where $M$ is the nucleus mass, $E_\nu$ is the energy of the incoming neutrino. $T_{\rm max}$ is the maximal value of
recoil energy, $T_{\rm max}=\frac{2E_\nu^2}{M+2E_\nu}$. The constants $\xi_{S,V,T}$ and $A_M$ in the above equation represent the effective parameters characterizing neutrino-nucleus interactions mediated by scalar, vector, tensor, and dipole currents, respectively. By comparing Eqs.~\ref{eq:gV}-\ref{eq:gT} and \ref{eq:ds}, one can directly establish a correspondence between the Wilson coefficients and the $\xi/A_M$ parametrization. Assuming that only one effective parameter is present at a time, constraints on these parameters were derived by fitting the COHERENT data in Refs.~\cite{AristizabalSierra:2018eqm,Giunti:2019xpr,Chang:2020jwl}. From the above differential equation we can identify the
dipole operator contribution as,
\begin{align}
A_M^2=\mathbb{Z}_i^2\sum_{\beta}\Big|\frac{1}{G_F}\mu_{\nu N}^{\alpha\beta}\Big|^2 F_p(q^2).    
\end{align}
Using the result in Ref.~\cite{Chang:2020jwl}, the $90\%$ CL bound on the dipole operators is given by
\begin{align}
\sum_\beta\Big|\frac{1}{G_F v}\mu_{\nu N}^{\alpha\beta}\Big|^2 < 1.44\times 10^{-7},    
\label{eq:mm}
\end{align}
where we sum over the final state neutrino flavor.
Similarly following Ref.~\cite{AristizabalSierra:2018eqm}, the $90\%$ CL bounds for the $\xi_S$ and $\xi_T$ parameters are given as,
\begin{align}
\label{eq:xi_s}
& {\xi_S^2\over \mathbb{N}^2F^2}
=
\sum_{\beta,i}\Big| \sum_{q=u,d} 2(\tilde{g}_S^{qq})_{\alpha\beta}\left({\mathbb{Z}_i\over \mathbb{N}_i}{m_p\over m_q}f^p_{T_q}+{m_n\over m_q}f^n_{T_q}\right)\Big|^2<0.62^2 \, , \\
& {\xi_T^2\over \mathbb{N}^2F^2}
=8\sum_{\beta,i}\Big| \sum_{q=u,d} 2 (\tilde{g}_T^{qq})_{\alpha\beta}\left({\mathbb{Z}_i\over \mathbb{N}_i}\delta_q^p+\delta_q^n\right)\Big|^2<0.591^2 \;.
\label{eq:xi_t}
\end{align}
These bounds apply for initial state neutrino flavor $\alpha=e$ or $\mu$. Assuming all the Wilson coefficients which enter in Eqs.~\ref{eq:mm}, \ref{eq:xi_s} and \ref{eq:xi_t} to be equal and $C \sim 1$, these give the bound on the cutoff scale as $\Lambda> 27$ TeV, 1.9 TeV and 220 GeV, respectively.
\black
\section{Constraints from stellar cooling and Supernova 1987A}
\label{sec:stellar-colling}
RHNs do not interact via the standard weak interaction and hence can escape the star unhindered if they are produced in the hot cores of stars. In that case, they would act as an effective energy sink, significantly accelerating the rate at which the star loses energy. The primary possibilities include the existence of new right-handed interactions that directly couple to right-handed neutrinos, as well as the presence of neutrino magnetic or electric dipole moments, which would enable left-right scatterings~\cite{Raffelt:1999gv,Raffelt:1999tx,Raffelt:1996wa,Raffelt:1990yz}. Among the operators listed in Table.~\ref{tab:bosonic} and Table.~\ref{tab:fermionic}, there are many which can enable the efficient $N_R$ production.
\par As we already discussed, the magnetic moments $\mu_{\nu N}$ results from the combination of $C_{LNB}\mathcal{O}_{LNB}+C_{LNW}\mathcal{O}_{LNW}$ and in presence of nonzero magnetic moment, the electromagnetic excitations~(called plasmons) inside the hot core plasma gives rise to two body decays into $\nu_L+N_R$. 
Thus, an additional cooling by escaping $N_R$  can impose a stringent
upper limit on $\mu_{\nu N}$. We use the recent analysis of global clusters from Refs.~\cite{Raffelt:1999gv,Raffelt:1996wa}, which sets an upper
limit on the neutrino magnetic moment as $\mu_{\nu N}< 3\times 10^{-12}\mu_B$. This further implies a lower bound on the cutoff scale as $\Lambda > 6.8\times 10^5$ GeV, assuming the coefficient to be $C_{LNB,LNW}\sim 1$.
\par In presence of right-handed charged current interaction such as $\mathcal{O}_{duNe}$, a supernova~(SN) core can loses energy into $N_R$ states as an ``invisible channel" by the process $e^-p\to N_R n$~\cite{Barbieri:1988av} with the following cross section
\begin{align}
\sigma(e^-p\to N_R n)\approx 4\epsilon_{\rm CC}^2 G_F^2 \frac{E_e^2}{\pi},\,\, \text{with  } \epsilon_{\rm CC}=\frac{\sqrt{2}}{4\Lambda^2}\frac{C_{duNe}}{G_F}.
\end{align}
The SN 1987A energy-loss argument then requires $\epsilon_{\rm CC}<10^{-5}$~\cite{Raffelt:1996wa,Raffelt:1987yt}. This translates to a lower bound on the cutoff scale as $\Lambda>55$~TeV assuming the coefficient to be $C_{duNe}\sim 1$. On the other hand, in the presence of right-handed neutral currents, the process such as $NN\to NN + N_R\bar{N}_R$ might play an important role in RHNs emission. Following Ref.~\cite{Raffelt:1996wa}, if we parametrized the relevant neutral current interaction as 
\begin{align}
\mathcal{L}_{\rm NC}=\frac{4\epsilon_{\rm NC} G_F}{\sqrt{2}}\overline{\psi_f}\gamma_\mu P_{R,L}\psi_f \,\overline{\psi_{f'}}\gamma_\mu P_{R}\psi_{f'},
\end{align}
then one obtains the bound $\epsilon_{\rm NC}< 3\times 10^{-3}$~\cite{Raffelt:1996wa}.
The operators such as $\mathcal{O}_{uuNN,ddNN}$ and $\mathcal{O}_{QQNN}$ contributes to this process and accordingly the parameter $\epsilon_{\rm NC}$ defined as,
\begin{align}
\epsilon_{\rm NC}^2=\left(\frac{\sqrt{2}}{4G_F\Lambda^2}\right)^2 \left(\left(C_{uuNN}+C_{ddNN}\right)^2+4C_{QQNN}^2\right).
\end{align}
This translates to the following bound on the cutoff scale $\Lambda>5.3$ TeV, assuming all the involved coefficients to be of order one.
In the Dirac case, the chirality violating operators such as $\mathcal{O}_{LdQN,LNQd}$ and $\mathcal{O}_{QuNL}$ will lead to a helicity flipping interaction of the Dirac neutrinos on nucleons~\cite{Thomas:1992hf,Gandhi:1990bq}. The relevant cross section for this process is given as follows
\begin{align}
\sigma(\nu_L N\to N_R N)\approx \frac{A}{32\pi}\frac{m_{N}^2}{\Lambda^4 m_u^2 m_d^2}E_\nu^2,\,\, \text{with  } A=m_u f_d^N\left(C_{LdQN}-2C_{LNQd}\right) - 2 m_d f_u^N C_{QuNL}),
\end{align}
where $f_q^N=m_q\braket{N|\bar{q}q|N}/2m_N^2$~\cite{Borsanyi:2020bpd}. References.~\cite{Gandhi:1990bq,Grifols:1990jn} determine a bound on the helicity-flipping cross section of $\sigma < 2.4\times 10^{-48}\,\text{cm}^2$ by imposing the condition that the observed neutrino pulse is not significantly shortened by such cooling. Then, taking 30 MeV as the average neutrino temperature in the SN core gives the bound on the cutoff scale as $\Lambda>24$ TeV, where we assumed the relevant coefficient to be all equal and of order one. In presence of four-fermion operators such as $\mathcal{O}_{eeNN}$ and $\mathcal{O}_{LLNN}$, $N_R\bar{N}_R$ emission can take place via the process $e^+e^-\to N_R\bar{N}_R$~\cite{Barbieri:1988av}. The relevant cross section is given as,
\begin{align}
\sigma(e^+e^-\to N_R\bar{N}_R)=\frac{(C_{eeNN}^2+C_{LLNN}^2)}{48\pi\Lambda^4}\,s,
\end{align}
where $s$ is the center-of-mass energy squared. Again similar considerations as before lead to bounds on the cutoff scale $\Lambda>1.8$ TeV if we assume the SN inner core temperature as $T_C=30$ MeV. In Table.~\ref{tab:stellar-cooling}, we summarized the resulting bounds on the cutoff scale coming from the various stellar cooling processes. 
\renewcommand{\arraystretch}{1.2}
\begin{table}[ht]
\begin{tabular}{ ||c | c|c  || }
\hline 
\hspace{0.5cm} Process & \hspace{0.05cm} Relevant Coefficients \hspace{0.05cm}  & $\Lambda_{\rm min}$ [TeV]\\ 
\hline \hline
$\gamma\to \nu_L + N_R$   & $C_{LNB}$, $C_{LNW}$ & 680
\\
\hline
$e^-p\to n + N_R$   & $C_{duNe}$ & 55\\
\hline
$NN\to NN + N_R\bar{N}_R$  & $C_{uuNN,ddNN}$, $C_{QQNN}$ & 5.3\\
\hline 
$\nu_L N\to N_R N$   & $C_{LdQN,LNQd}$, $C_{QuNL}$ & 24\\
\hline
$e^+e^-\to N_R + \bar{N}_R$   & $C_{eeNN,LLNN}$ & 1.8
\\
\hline
\end{tabular}
\caption{Constraint on the cutoff scale coming from the stellar cooling, assuming all the involved Wilson coefficients are of order one.}
\label{tab:stellar-cooling}
\end{table} 
\par Note that the above discussion is by no means complete, and a more systematic study is needed considering all the processes by which RHNs can be produced in the hot cores of stars. We leave this for future work.
\begin{figure}[b!]
    \centering
    \includegraphics[scale=0.4]{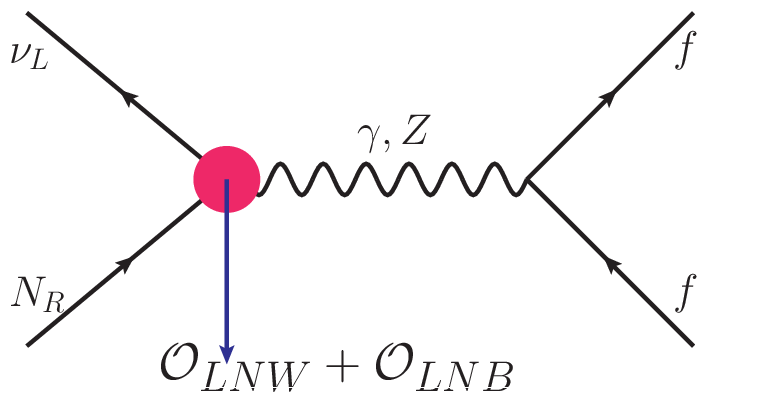}\,
    \includegraphics[scale=0.4]{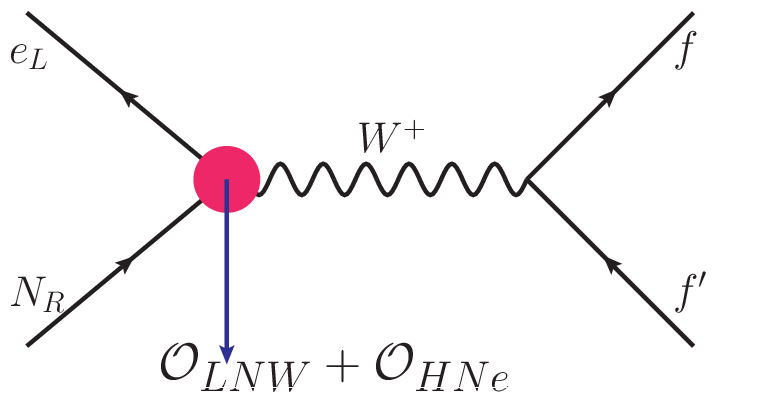}\,
    \includegraphics[scale=0.4]{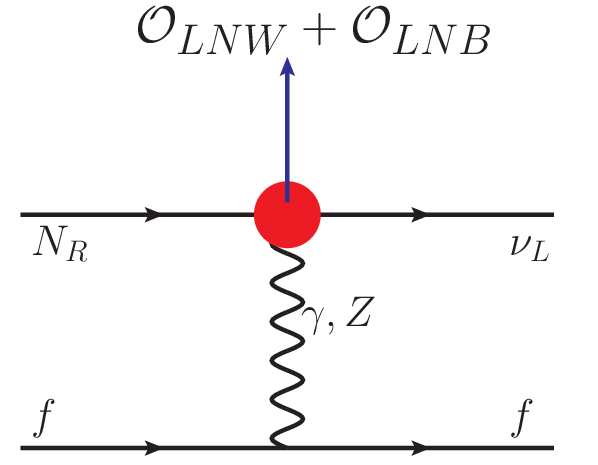}\\
    \includegraphics[scale=0.4]{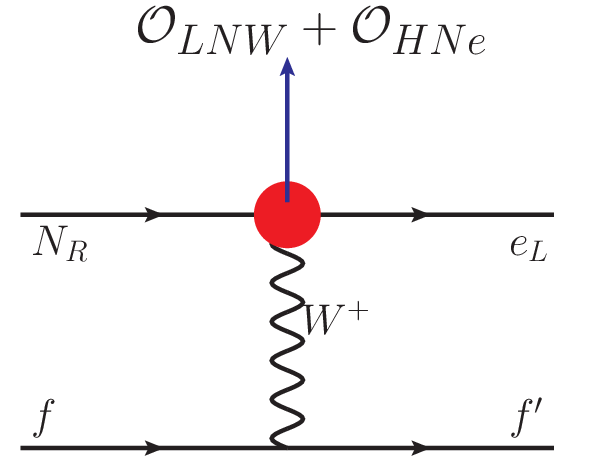}\,
    \includegraphics[scale=0.4]{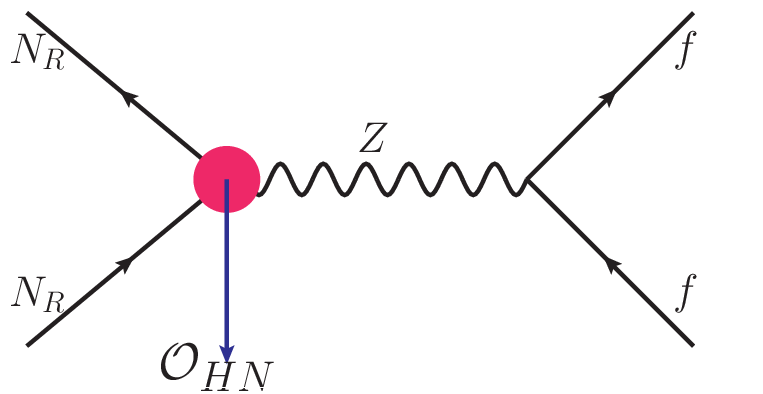}
    \caption{Feynman diagrams that are responsible for thermalizing the $N_R$ in the early Universe.}
    \label{fig:NRtherm}
\end{figure}
\section{Cosmological Signatures: imprints on $\Delta{N_{\rm eff}}$}
\label{sec:Neff}
The anisotropic behavior in the remnant radiation from the early Universe,
known as the cosmic microwave background radiation (CMBR), is extremely
sensitive to the presence of any extra radiation energy at the time
of recombination ($z\approx 1100$) \cite{Jungman:1995bz}. The amount of
radiation energy density that was present in the early Universe, except for the contribution coming from photons, is usually parametrized in terms of the
effective numbers of neutrino species \cite{Mangano:2005cc},
\begin{eqnarray}
    N_{\rm eff} \equiv \frac{\rho_{R} - \rho_{\gamma}}{\rho_{\nu_L}},
    \label{eq:Neff}
\end{eqnarray}
where $\rho_R$ is the total radiation energy density, $\rho_\gamma$ is the energy density of photon, and $\rho_{\nu_L}$ is the energy density of a single
active neutrino species. The latest measurements of the cosmic microwave
background (CMB) from the Planck satellite \cite{Planck:2018vyg}, combined
with baryon acoustic oscillation (BAO) data indicate that
$N_{\rm eff} = 2.99^{+0.34}_{-0.33}$ at $95\%$ CL which perfectly agrees
with the SM prediction $N^{SM}_{\rm eff} = 3.0440\pm 0.0002$ \cite{Mangano:2005cc, Grohs:2015tfy,deSalas:2016ztq, Froustey:2020mcq, Bennett:2020zkv, Akita:2020szl}. On the
other hand, the next generation CMB experiments, for example, CMB-S4 \cite{Abazajian:2019eic}, SPT-3G \cite{SPT-3G:2019sok}, LiteBIRD \cite{LiteBIRD:2022cnt}, are going to improve their sensitivity
and planning to probe $\Delta{N_{\rm eff}} = N_{\rm eff} - N_{\rm eff}^{SM}
= 0.06$ at $95\%$ CL. Such precise measurement of $\Delta{N_{\rm eff}}$
is expected to test the presence of light degrees of freedom (DOF) that
were either in equilibrium with the SM particles at some point of time during the evolution of our Universe or yielded from the nonthermal decay and
annihilation of bath particles \cite{Abazajian:2019oqj, FileviezPerez:2019cyn,Adshead:2020ekg,Luo:2020sho,Luo:2020fdt,Han:2020oet,Li:2021okx,Okada:2022cby,Chen:2015dka}. Here, we have considered neutrinos are Dirac fermions
and there are three right-handed neutrinos $(N_R)$ that have interactions with
the SM particles at dimension six level as shown in Table \ref{tab:bosonic} and Table \ref{tab:fermionic}. If these three $N_R$s were present in the thermal bath of the early Universe, the data from Planck 2018 suggests that they must decouple
from the SM plasma at a much higher temperature than the SM neutrinos,
around 600 MeV \cite{Xing:2020ijf}. Otherwise, their effect on
$\Delta{N_{\rm eff}}$ will be more than the currently allowed
limit of 0.286 at $95\%$ CL \cite{Planck:2018vyg}. As a matter of fact,
the limit on decoupling temperature also determines the bounds on
the interactions of $N_R$.  If the neutrino masses are generated solely
through the standard Higgs mechanism, the contribution of $N_R$ to
$N_{\rm eff}$ would be negligible $\left(\mathcal{O}(10^{-12})\right)$ \cite{Luo:2020fdt}. This is due to the extremely small Yukawa
couplings, which prevent $N_R$ from reaching thermal equilibrium
with the SM plasma. Therefore, any future cosmological detection of
Dirac neutrinos in upcoming CMB experiments would imply the existence
of new interactions in the neutrino sector.   

In Table \ref{tab:bosonic} and Table \ref{tab:fermionic}, we have listed
all possible dimension six operators through which $N_R$ can interact
with the SM particles and become thermalized in the early Universe. In \cite{Luo:2020sho,Biswas:2022fga}, the upper limits on the four-fermion
interactions between $\nu_L$ and $N_R$ were discussed by considering the
impact of new physics in $\Delta{N_{\rm eff}}$ in the context of standard
and nonstandard cosmological expansion history. This is because, at 600 MeV or higher temperature, all the light fermions will behave similarly and will not change the bounds drastically. Here, we will mostly focus on the bounds of the operators given in Table \ref{tab:bosonic} as they will open up new interactions of $N_R$ that were not discussed earlier.
From Eq.~\eqref{eq:Neff}, the additional contribution to $N_{\rm eff}$ coming due to the presence of relativistic $N_R$ at the time of recombination can be written as 
\begin{eqnarray}
    \Delta{N_{\rm eff}} = \frac{\sum_{\alpha} \rho_{N_R}^\alpha}{\rho_{\nu_L}} 
    = 3\times \frac{\rho_{N_R}}{\rho_{\nu_L}} 
    = 3\times \left(\frac{T_{N_R}}{T_{\nu_L}} \right)^4,
    \label{eq:DeltaNeff}
\end{eqnarray}
where $\alpha$ represents the number of the generations of $N_R$ in the theory and we are taking $\alpha = 3$. During the derivation of the above equation, we also assume that all three $N_R$s have identical interactions with the SM, and as a result, the total energy density of $N_R$ can be evaluated by multiplying the energy density of a single $N_R$ by 3 as $\sum_{\alpha} \rho_{N_R} = 3 \times \rho_{N_R}$. To estimate $\Delta{N_{\rm eff}}$, one needs to track the temperature of $N_R$ ($T_{N_R}$), which evolves independently of the bath temperature after the decoupling. Once $N_R$ decouples from the thermal bath, the energy density $\rho_{N_R}$ changes only because of the expansion and redshifts as $a(t)^{-4}$. The energy density of the active neutrinos also scales similarly before and after the electron-positron annihilation. Neutrinos decouple from the thermal bath before $e^\pm$ annihilation and remain unaffected by the electron-positron annihilation, which increases the photon temperature than the active neutrinos. So, practically, there is no need to track $T_{N_R}$ all the way up to CMB; rather, it is sufficient to evaluate the ratio at a higher temperature $T\, \left(T>T_{\nu_L}^{dec} >> T_{\rm CMB}\right)$. Let us define the ratio\footnote{We have solved $\xi_{N_R}$ well before
the decoupling of left handed neutrinos ($\sim 1$ MeV) and therefore
in that era $T_{\nu_L}=T$, where $T$ is the temperature of thermal bath
comprised of the SM particles.} of two temperatures $\xi_{N_R} \equiv \frac{T_{N_R}}{T_{\nu_L}}$. Now, to track the evolution of $\xi_{N_R}$
(equivalently $T_R$), we need to solve the Boltzmann equation for the energy density of $N_R$, which can be expressed in terms of $\xi_{N_R}$ as shown in \cite{Biswas:2021kio}
\begin{eqnarray}
\dot{\rho}_{N_R} + 3 H \left(\rho_{N_R}+P_{N_R} \right) &=& \mathcal{C}_{N_R},\\
T\dfrac{\,d\xi_{N_R}}{dT} + \left(1 - \beta \right) \xi_{N_R} &=& -\dfrac{\beta}
{4\,\kappa\,\xi_{N_R}^3 H\,T^4} \mathcal{C}_{N_R},
\label{eq:BETNR}
\end{eqnarray}
where $\beta = \dfrac{g_{\star}^{1/2}(T)\sqrt{g_{\rho}(T)}}{g_s(T)}$ with $g_{\rho}$ and $g_s$ are the effective DOFs associated with energy density and entropy density respectively while $g_{\star}^{1/2}=\dfrac{g_s}{\sqrt{g_{\rho}}}\left(1 + \dfrac{1}{3}\dfrac{T}{g_s} \dfrac{dg_s}{dT}  \right)$ and $\kappa = 2\times \dfrac{7}{8} \dfrac{\pi^2}{30}$. The detailed derivation of the collision term $\mathcal{C}_{N_R}$ is given in Appendix \ref{BEeq}. The relevant Feynman diagrams that are responsible for thermalizing $N_R$ in the early Universe are shown in Figure~\ref{fig:NRtherm}, and the analytical expressions for cross sections $(\sigma_{N_{R}+X \rightarrow Y+Z})$ of different processes are given in Appendix.~\ref{appen:Xsec}. One can note that the processes can be mediated by the SM gauge bosons, and as a result, the new physics appears only on one side of the Feynman diagrams. 
\begin{figure}[t!]
\includegraphics[scale=0.45]{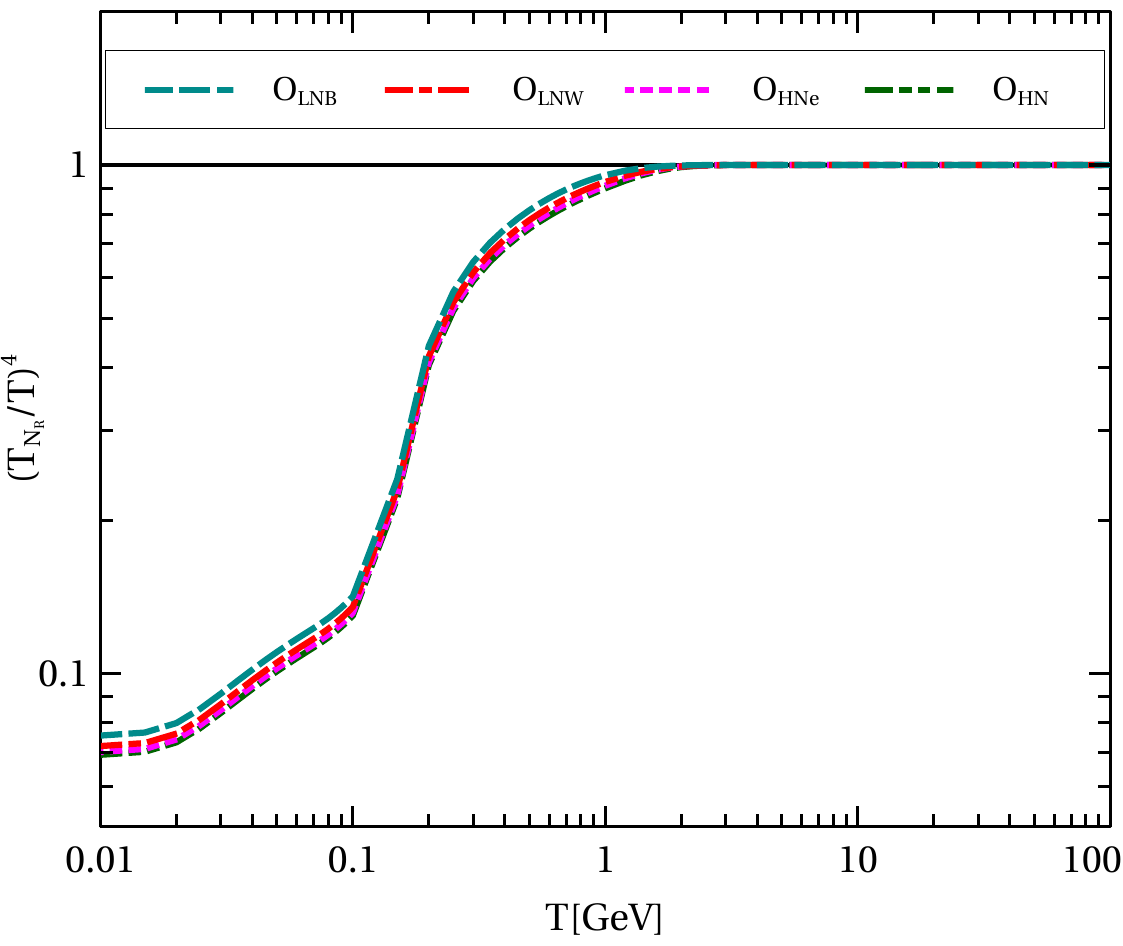}
\caption{The evolution of $T_{N_R}/T$ for different operators. The cutoff scale is taken to be $\Lambda=10^5$ GeV}
\label{fig:xi}
\end{figure}
\begin{figure}[t!]
\includegraphics[scale=0.5]{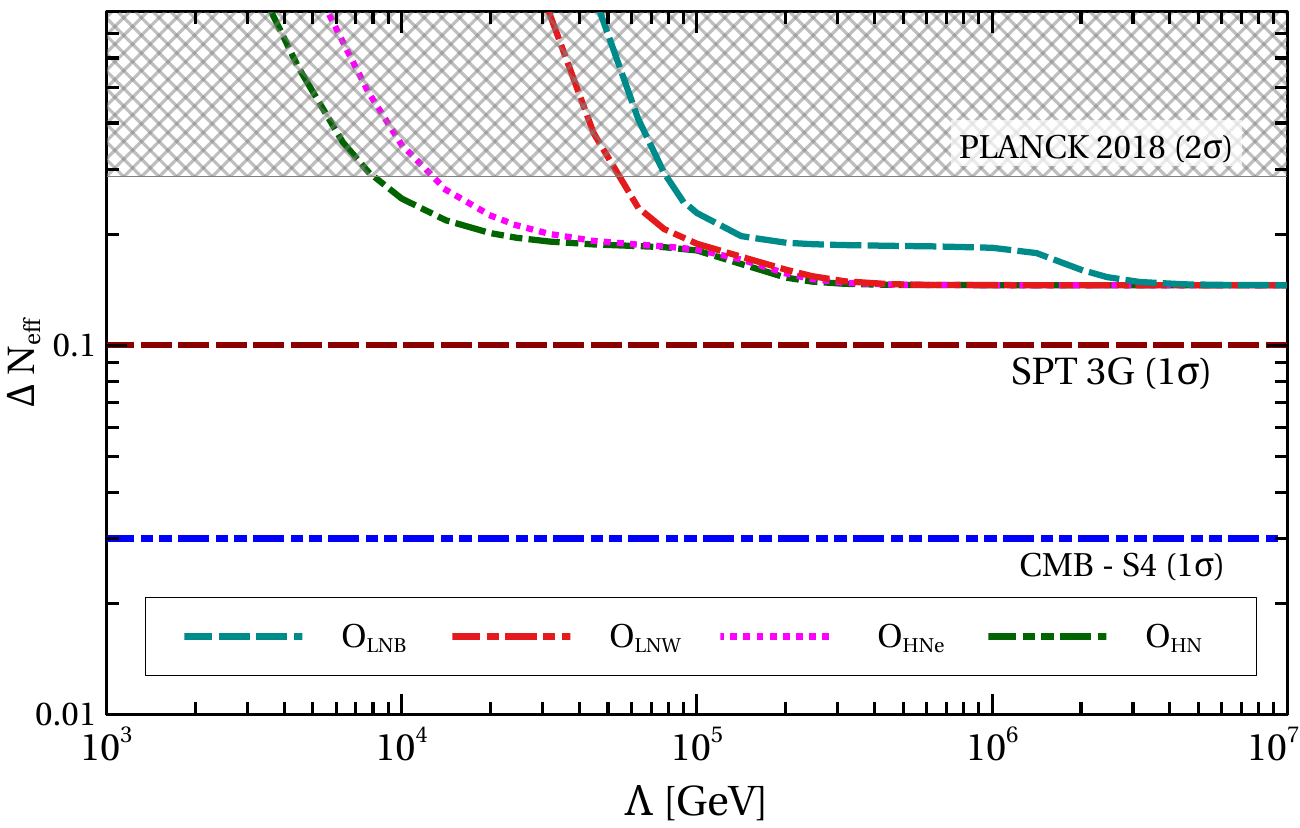}
\caption{The contribution to $\Delta{N_{\rm eff}}$ as a function of cutoff scale coming from the presence of $N_R$ in the thermal plasma in the early Universe. }
\label{fig:Neff}
\end{figure}
\begin{table}[b!]
    \centering
    \begin{tabular}{||c|c||}
 \hline
 cutoff scale 
 & Constraint from $N_{\rm eff}$ \\ 
 & (TeV) \\
 \hline\hline
 $\Lambda_{\rm LNB}$ & $80$\\ 
 \hline
 $\Lambda_{\rm LNW}$ & $50$\\
 \hline
 $\Lambda_{\rm HNe}$ &  $15$\\
 \hline
 $\Lambda_{\rm HN}$ & $8$\\
 \hline
\end{tabular}
    \caption{Constraints on different cutoff scales from their contribution to $\Delta{N_{\rm eff}}$.}
    \label{tab:Neff}
\end{table}
By solving Eq.~\eqref{eq:BETNR} we have found $\xi_R$ as a function of SM plasma temperature $(T)$ and evaluated the contribution to $\Delta{N_{\rm eff}}$ by using Eq.~\eqref{eq:DeltaNeff}. In Figure \ref{fig:xi}, we have shown the evolution of $\xi_{N_R}$ as a function of $T$ where different lines correspond to different operators. Initially, $N_R$ was in the thermal plasma having the same temperature as SM particles. Once it decouples from the bath, its temperature evolves independently. Here, for all the operators, the cutoff scale has been fixed as $\Lambda = 10^{5}$ GeV. The value of  $\Delta{N_{\rm eff}}$ as a function of the mass scale $(\Lambda)$ coming from the presence of $N_R$ in the thermal plasma in the early Universe is shown in Figure~\ref{fig:Neff} where the hatched region represents the region that is already excluded by Planck 2018 data at 2$\sigma$ CL The two horizontal lines, brown dashed line, and blue dot-dot-dashed line, show the future predictions of SPT-3G and CMB-S4 at 1$\sigma$ CL The green, magenta, red, and cyan colored lines correspond to the contribution coming from the operators $\mathcal{O}_{HN}$, $\mathcal{O}_{HNe}$, $\mathcal{O}_{LNW}$, and $\mathcal{O}_{LNB}$ respectively. One important point to note here is that in our calculation, we have considered only one operator at a time by keeping all the other operators to zero. The reason is as follows, the final contribution to $N_{\rm eff}$ depends on the decoupling temperature of $N_R$ from the thermal plasma, and that will be decided by the strongest one. In our calculation, we focus on only one operator at a time, setting all others to zero. This approach is based on the fact that the final contribution to $N_{\rm eff}$ is controlled by the decoupling temperature of $N_R$ from the thermal plasma. The decoupling temperature, in turn, is determined by the strongest operator in effect. Consequently, even if we were to allow multiple operators to be nonzero, the result would remain unchanged, provided these operators are nondegenerate, meaning that one operator's effect dominates due to a higher strength or efficiency at controlling the decoupling temperature. The most stringent bound is coming from the operator $\mathcal{O}_{LNB}$ as it involves the electromagnetic interaction. The Planck 2018 data already excludes $\Lambda_{LNB}\approx 80\text{ TeV}$, $\Lambda_{LNW}\approx 50\text{ TeV}$, $\Lambda_{LHNe}\approx 15 \text{ TeV}$, and $\Lambda_{LHN}\approx 8\text{ TeV}$ as shown in Table \ref{tab:Neff}. The hierarchy among the different cutoff scales can be understood as follows. In the case of $\mathcal{O}_{LNB}$ and $\mathcal{O}_{LNW}$, the photon-mediated diagram dominates the cross section, whereas for $\mathcal{O}_{HNe}$ and $\mathcal{O}_{HN}$, it is mediated by W and Z bosons respectively. That is why the bounds on the cutoff scale $\Lambda_{LNB}$ and $\Lambda_{LNW}$ are almost one order of magnitude stronger than $\Lambda_{HNe}$ and $\Lambda_{HN}$. In addition, in the case of $\mathcal{O}_{LNB}$, the photon couples to $N_R$ with the {\it cosine} of the {\it Weinberg angle} ($\theta_W$), making the coupling stronger than $\mathcal{O}_{LNW}$ where it comes with the {\it sine} of $\theta_W$. On the other hand, in the case of $\mathcal{O}_{HNe}$, there are two diagrams mediated by the $W$ boson, whereas, in the case of $\mathcal{O}_{HN}$, the only annihilation channel is mediated by the $Z$ boson. Also, there is an extra $\sqrt{2}$ factor present in the vertex with $W$ that makes the bound on $\Lambda_{HNe}$ a little stronger than $\Lambda_{HN}$. Different vertex factors, along with the involved fields, have been shown in Table \ref{tab:Vertex}. In the case of the operators discussed in Table \ref{tab:fermionic}, $\Delta{N_{\rm eff}}$ can exclude the operators such as $\Lambda_{eeNN}$ up to 12 {TeV} \cite{Luo:2020sho}. The predictions from future observations, such as CMB-S4 \cite{Abazajian:2019eic}, SPT-3G \cite{SPT-3G:2019sok}, brown and blue horizontal lines, show that they can exclude the full parameter space. {Meaning, there will always be a minimum contribution to $\Delta{N_{\rm eff}}$ that does not depend on the decoupling temperature. For three $N_R$s, the minimum contribution to $\Delta{N_{\rm eff}}$ will be $\approx 0.14$ \cite{Abazajian:2019oqj}.} In other words, future CMB observations will be capable of excluding the possibility of the presence of three ultralight right-handed neutrinos in the thermal plasma of the early Universe.  
\section{Concluding Remarks}
\label{sec:conclusion}
In this paper, we have studied the phenomenology of the dimension six operators in the context of $\nu$SMEFT, considering that the neutrinos are Dirac in nature. We have derived constraints on the cutoff scale $\Lambda$ by analyzing various high and low energy observables, assuming the involving Wilson coefficients of the order one. 
\par For this goal, we have relied on data from LHC searches for $\ell+\slashed{E}_T$ and $j+\slashed{E}_T$; on measurements of different proton, pion, tau, and top decays; leptonic and invisible decays of mesons; neutrino-nucleus scattering; beta decays as well as on the effect of extra Dirac states on the $N_{\rm eff}$. In the case of $\ell+\slashed{E}_T$ and $j+\slashed{E}_T$, our limits on the cutoff scale range from 1 TeV to 3 TeV for couplings of order $\sim 1$. We find that the upper bound on the proton lifetime translates into a lower bound on the cutoff scale $\Lambda$ of about $10^{15}$ GeV. The decay mode $p\to\pi^0 e^+$ provides the tightest bound as this mode is more severely restricted than other modes. We find that observation/nonobservation of particular proton or neutron decay modes might hint at the existence of RHNs. The bounds from various meson decays range from 700 GeV to 2.4 TeV, where the tightest bound comes from the decay mode $\pi^+\to\mu^+\nu_\mu$ as this is precisely measured. Tau leptonic and semileptonic decay width measurements can also give competitive bounds on the cutoff scale, which are of the order of 1 TeV. Unlike the meson or tau decay width measurements, the bound from top decay mode $t\to\ell_\alpha b+\slashed{E}_T$ is rather loose as the current bounds on the top width are not constraining enough. Instead, one can get a tighter bound~($\Lambda\sim 2$ TeV) using the $t\bar{t}$ production at LHC, where one of the tops decays leptonically through the modified dimension six vertex, while the
other tops decay identically as in the SM in the hadronic mode. We find that strong constraints on the EFT operators might arise from the $N_R$ production in the hot cores of stars and the effects of this
on stellar cooling. This gives a constraint on the cutoff scale $\Lambda$ in the ball park of 10 TeV. Finally, the contribution to $N_{\rm eff}$ coming due to the presence of three thermalized RHN sets a lower bound on $\Lambda$ from $8$ TeV to $80$ TeV, where the constraint on $\Lambda_{LNB}$ is the strongest. 
\par Finally we would like to mention that although some of these bounds are already exists in the literature, our study include many novel results such as: we classify EFT operators that contribute to the Dirac neutrino mass due to the spontaneous breaking of chiral symmetry via the light quark condensate and discuss
the resulting bounds; detailed discussion on the interplay of EFT operators to proton and neutron decay; systematic study of all relevant dimension six operator’s contribution to stellar cooling
and CMB radiation.
\black
\section*{Note Added}
During the final stages of this work, Ref.~\cite{Borah:2024twm} appeared, in which the bounds on some operators coming from $\Delta N_{\rm eff}$ are derived, which agrees with our results.
\section*{Acknowledgements}
The work of S.M. is supported by KIAS Individual Grants (PG086002) at Korea Institute for Advanced Study.
\appendix
\section{Details of the Proton/neutron decay calculation}
\label{app:proton_decay}
The baryon number violation but preserving $B-L$ number is generally expressed as low-energy effective Hamiltonian with the above six-dimension operators. The relevant effective Lagrangian can be written as~\cite{Aoki:2013yxa,Yoo:2021gql,JLQCD:1999dld},
\begin{align}
\mathcal{L}^{\slashed{B}} = \sum_I C^I \Big[(qq)(q\ell)\Big]^I = - \sum_I C^I \Big[\overline{\ell^c} \mathcal{O}_{qqq}\Big]^I,
\end{align}
where $C^I=C^I(\mu)$ is the Wilson coefficient with renormalization scale $\mu$ of the corresponding operators with $q$ being a light quark flavor $u$, $d$, or $s$. The details of the UV completed model are all captured in the Wilson coefficients $C^I(\mu)$. The three quark operator reads as
\begin{align}
\mathcal{O}_{qqq}^{\chi\chi'}=(qq)_{\chi} q_{\chi'}=\epsilon^{\alpha\beta\gamma} (q^{\alpha T}C P_{\chi} q^\beta) P_{\chi'} q^\gamma
\end{align}
where the color singlet contraction is taken, $\{\bar{q},\bar{\ell}\}^C=\{q,\ell\}^T C$ are charge-conjugated fields and the chirality projectors $P_{R,L}=\frac{1\pm\gamma_5}{2}$. From now on, we use simple notations for the three-quark operators as $\mathcal{O}^{\chi\chi'}$ where $\chi$ and $\chi'$ denote the chirality, either $R$ or $L$. Now we are ready to calculate the transition matrix elements of the BNV dimension-six operators with an initial nucleon~(proton or neutron, $N = p, n$) state and a final state containing a pseudoscalar meson~($P = (\pi, K, \eta)$) and an antilepton ($\bar{\ell}$),
\begin{align}
\bra{P(\vec{p}),\bar{\ell}(\vec{q},s)}\overline{\ell^c}\mathcal{O}^{\chi\chi'}\ket{N(\vec{k})}=\overline{v_{\ell}^c}(\vec{q},s)\bra{P(\vec{p})} \mathcal{O}^{\chi\chi'}\ket{N(\vec{k},s)},
\label{eq:current}
\end{align}
where $\vec{k}$ and $\vec{p}$ stand for the three momentum of the initial state nucleon and the final state pseudoscalar meson, $\vec{q} = \vec{p} - \vec{k}$ for the final lepton. Leptonic matrix element can be written as $\bra{\ell(\vec{q},s)} \overline{\ell^c}\ket{0}=\overline{v_\ell^c}(\vec{q},s)$ whereas the hadronic part $\bra{P(\vec{p})} \mathcal{O}^{\chi\chi'}\ket{N(\vec{k},s)}$ is parametrized by the relevant form factor $W_0(q^2)$ and irrelevant one $W_1(q^2)$ as,
\begin{align}
\bra{P(\vec{p})} \mathcal{O}^{\chi\chi'}\ket{N(\vec{k},s)}=P_{\chi'} \Big[W_0^{\chi\chi'}(Q^2)-\frac{i\slashed{q}}{m_N} W_1^{\chi\chi'}(Q^2)\Big] u_N(k,s)
\end{align}
The form factors $W_0$, and $W_1$ are defined for each matrix element with the three-quark operator renormalized in the $\overline{\text{MS}}$ scheme at the scale $\mu$. The form factors are functions of the square of four-momentum transfer $q=k-p$. Through the Parity transformation, the various chirality combinations of the matrix elements are connected as
\begin{eqnarray}
  &&\langle P(\vec p)|\mathcal O^{R L}|N(\vec k,s)\rangle
  = \gamma_0\langle P(-\vec p)|\mathcal O^{L R}|N(-\vec k,s)\rangle,\\
  &&\langle P(\vec p)|\mathcal O^{L L}|N(\vec k,s)\rangle
  = \gamma_0\langle P(-\vec p)|\mathcal O^{R R}|N(-\vec k,s)\rangle.
\end{eqnarray}
The consequence of Parity transformation is that four chirality combinations $(\chi\chi')=(RL),(LL),(LR),(RR)$ are reduced to two different combinations, $(\chi\chi')=(RL),(LL)$. As a result, in the following $\chi'$ is fixed in a left-handed chirality, and a short-hand notation $W_{0,1}^{\chi L}\equiv W_{0,1}^{\chi}$ will be used. Under the ``isospin symmetry"~(exchange-symmetry between $u$ and $d$) there are the following relations between proton and neutron matrix elements~\cite{Aoki:2006ib}:
\begin{eqnarray}
  \langle \pi^0| (ud)_\chi u_L | p\rangle &=& \langle \pi^0| (du)_\chi d_L | n\rangle,
  \label{eq:n_p_1}\\
  \langle \pi^+| (ud)_\chi d_L | p\rangle &=& -\langle \pi^-| (du)_\chi u_L | n\rangle,
  \label{eq:n_p_2}\\
  \langle K^0| (us)_\chi u_L | p\rangle &=& -\langle K^+| (ds)_\chi
   d_L | n\rangle, \label{eq:n_K0}\\
  \langle K^+| (us)_\chi d_L | p\rangle &=& -\langle K^0| (ds)_\chi u_L | n\rangle,\\
  \langle K^+| (ud)_\chi s_L | p\rangle &=& -\langle K^0| (du)_\chi s_L | n\rangle,\\
  \langle K^+| (ds)_\chi u_L | p\rangle &=& -\langle K^0| (us)_\chi d_L | n\rangle,\\
  \langle \eta| (ud)_\chi u_L | p\rangle &=& -\langle \eta| (du)_\chi d_L | n\rangle.
  \label{eq:n_eta}
\end{eqnarray}
The negative sign is the artifact of the interpolation operator of the proton or neutral pion by the exchange of $u$ and $d$. In addition, isospin symmetry requires that,
\begin{equation}
  \langle \pi^0| (ud)_\chi u_L | p\rangle = \sqrt 2\langle \pi^+| (ud)_\chi d_L | p\rangle.
  \label{eq:piplus}
\end{equation}
 Using the on-shell condition, the relevant matrix element for the nucleon decay can be written as,
\begin{align}
\overline{v_\ell^c}(\vec{q},s)\bra{P(\vec{p})} \mathcal{O}^{\chi L}\ket{N(\vec{k},s)}& =\overline{v_\ell^c}(q,s) P_{L} \Big[W_0^{\chi}(Q^2)-\frac{i\slashed{q}}{m_N} W_1^{\chi}(Q^2)\Big] u_N(k,s) \nonumber \\
& = \overline{v_\ell^c}(q,s) P_{L} u_N(k,s) W_0^{\chi}(-m_\ell^2) + \mathcal{O}(m_\ell/m_N),
\label{eq:matrix-element}
\end{align}
where $Q^2=-q^2=-(E_N-E_P)^2+(\vec{k}-\vec{p})^2$. As $q^2 =m_{\ell}^2$ is much smaller than nucleon mass squared in the case of $\ell=e,\nu$, one can, with very good approximation, set $q^2\approx 0$ and drop the second term in Eq.~\eqref{eq:matrix-element}. This is why sometime in literature, $W_0$ is called relevant, and $W_1$ is irrelevant form factor. In this case with the knowledge of form factor $W_0$ from lattice QCD calulation, the partial decay width of $N\to P+\bar{\ell}$ can be estimated as,
\begin{align}
\Gamma(N\to P+\bar{\ell})=\frac{m_N}{32\pi\Lambda^4}\Big[1-\left(\frac{m_p}{m_N}\right)^2\Big]^2 \Big|\sum_I C^I W^I_0(N\to P)\Big|^2 ,
\label{eq:decay-rate}
\end{align}

\begin{table}[ht]
\begin{tabular}{ ||c | c | c  || }
\hline 
\hspace{0.5cm} Matrix element & \hspace{0.05cm} $W_0$ \hspace{0.05cm} & \hspace{0.05cm} $W_1$ \hspace{0.05cm}  \\ 
\hline \hline
$\bra{\pi^+} (ud)_L d_L \ket{p}$ & $0.1032 $ & $-0.130$    \\
                                 & $0.105$   & $-0.132$    \\
\hline 
$\bra{\pi^+} (ud)_R d_L \ket{p}$ & $-0.1125$ & 0.116       \\
                                 & $-0.1139$ & 0.118       \\
\hline
$\bra{K^0} (us)_L u_L \ket{p}$   & $0.0395$  & 0.0256      \\
                                 & $0.0397$  & 0.0254     \\
\hline
$\bra{K^0} (us)_R u_L \ket{p}$   & $0.0688$  & $-0.0250$   \\
                                 & $0.0693$  & $-0.0254$  \\
\hline
$\bra{K^+} (us)_L d_L \ket{p}$   & $0.0263$  & $-0.0448$   \\    
                                 & $0.0266$  & $-0.0453$   \\ 
\hline                                                                                                                                                                      
\end{tabular}\,
\begin{tabular}{ ||c | c | c  || }
\hline 
\hspace{0.5cm} Matrix element & \hspace{0.05cm} $W_0$ \hspace{0.05cm} & \hspace{0.05cm} $W_1$ \hspace{0.05cm}  \\ 
\hline \hline
$\bra{K^+} (us)_R d_L \ket{p}$   & $-0.0301$  & $0.0452$   \\
                                 & $-0.0307$  & $0.0458$   \\
\hline
$\bra{K^+} (ud)_L s_L \ket{p}$   & $0.0923$  & $-0.0638$   \\
                                 & $0.0932$  & $-0.0653$   \\
\hline
$\bra{K^+} (ud)_R s_L \ket{p}$   & $-0.0835$  & $0.0588$   \\    
                                 & $-0.0846$  &  $0.0605$ \\
\hline
$\bra{K^+} (ds)_L u_L \ket{p}$   & $-0.0651$  & $0.0192$ \\ 
                                 & $-0.0658$  & $0.0201$  \\                                
\hline  
$\bra{K^+} (ds)_R u_L \ket{p}$   & $-0.0394$  & $-0.0203$   \\         
                                 & $-0.0393$  & $-0.0204$ \\
\hline                                              \end{tabular}
\caption{Results for the form factors $W_{0,1}$ on the 24ID ensembles at the two kinematic points $Q^2=0$~(first line) and $Q^2 =-m_\mu^2$~(second line) renormalized to $\overline{\text{MS}}$~(2 GeV)~\cite{Yoo:2021gql}.}
\label{tab:matrix-element}
\end{table}
We have listed the relevant form factors $W_{0,1}$ for various matrix elements in Table.~\ref{tab:matrix-element}.
Note that both $C^I$ and $W^I_0(0)$ are renormalization scale dependent, but it cancels out in their multiplication. Note that although form factor $W_1$'s contribution can be disregarded for decays into positrons and antineutrinos, but not for those into antimuons as $m_{\ell}/m_N \approx 0.1$. In this case, the decay rate takes the following form~\cite{Yoo:2021gql},
\begin{equation}
\label{eqn:pdecay_rate_full}
\Gamma (N \rightarrow P \bar{\ell}  ) = \frac1{32\pi m_N}\lambda^{\frac{1}{2}}\left(1,\frac{m_\ell^2}{m_N^2},\frac{m_P^2}{m_N^2}\right)\left(\left(m_\ell^2+m_N^2-m_P^2\right)\left(|\tilde{W}_0|^2+\frac{m_\ell^2}{m_N^2}|\tilde{W}_1|^2\right)-4m_\ell^2\text{Im}\left(\tilde{W}_0 \tilde{W}_1^*\right)\right)\,,
\end{equation}
where $\tilde{W}_{0,1}$ are defined as,
\begin{align}
\tilde{W}_{0,1}=\sum_I \frac{C^I}{\Lambda^2} W^I_{0,1}(N\to P),
\end{align}
and $\lambda(x,y,z)=x^2+y^2+z^2-2xy-2xz-2yz$. Note that in the limit $m_{\ell}\to 0$ the decay rate Eq.~\eqref{eqn:pdecay_rate_full} is simplified to Eq.~\eqref{eq:decay-rate}.
\section{Details of the threebody leptonic decay mode: $\ell_\beta\to\ell_\alpha\nu N/\ell_\alpha N N$}
\label{app:tau-decays}
In the following, we give the details of the decay mode $\ell_\beta\to\ell_\alpha+\slashed{E}_T$. This decay mode can arise due to the following operators: $\mathcal{O}_{HNe/LNW}$, $\mathcal{O}_{LNLe}$ and $\mathcal{O}_{LLNN,eeNN}$. More specifically we can have following leptonic decay channels: $\ell_\beta^-\to\ell_\alpha^- N\overline{N}$, $\ell_\beta^-\to\ell_\alpha^-\nu\overline{N}$ and $\ell_\beta^-\to\ell_\alpha^-\overline{\nu}N$. The corresponding amplitude and spin averaged amplitude squared are given as follows:\\

1. $\ell_\beta^-(p)\to\ell_\alpha^-(k_1) N(k_2)\overline{N}(k_3)$:
\begin{align}
&\mathcal{M}(\ell_\beta^-\to\ell_\alpha^-N \overline{N}) = \frac{C_{eeNN}^{\alpha\beta NN}}{\Lambda^2}\, \overline{u}(\ell_\alpha)\gamma^\mu P_R u(\ell_\beta)\,\, \overline{u}(N)\gamma^\mu P_R v(\overline{N}) + \frac{C_{LLNN}^{\alpha\beta NN}}{\Lambda^2}\, \overline{u}(\ell_\alpha)\gamma^\mu P_L u(\ell_\beta)\,\, \overline{u}(N)\gamma^\mu P_R v(\overline{N}),\\
&\frac{1}{2}\sum |\mathcal{M}|^2  = \frac{8}{\Lambda^4}\Big\{\left(C_{eeNN}^{\alpha\beta NN}\right)^2(p.k_3) (k_1.k_2) + \left(C_{LLNN}^{\alpha\beta NN}\right)^2 (p.k_2) (k_1.k_3) - C_{eeNN}^{\alpha\beta NN} C_{LLNN}^{\alpha\beta N N} m_{\ell_\alpha} m_{\ell_\beta} (k_2.k_3)\Big\}.
\end{align}
2. $\ell_\beta^-(p)\to\ell_\alpha^-(k_1) \overline{N}(k_2)\nu_\beta(k_3)$:
\begin{align}
&\mathcal{M}(\ell_\beta^-\to\ell_\alpha^-\overline{N}\nu_\beta) =\frac{C_{HNe}^{N\alpha}}{\Lambda^2}\, \overline{u}(\nu_\beta)\gamma^\mu P_L u(\ell_\beta)\,\, \overline{u}(\ell_\alpha)\gamma_\mu P_R v(\overline{N}) - 2\frac{C_{LNW}^{\alpha N}}{M_W \Lambda^2}\, \overline{u}(\nu_\beta)\gamma_\mu P_L u(\ell_\beta)\,\, \overline{u}(\ell_\alpha)\sigma^{\nu\mu}P_R v(\overline{N}) (p_{\ell_\beta}-p_{\nu_\beta})_\nu\nonumber \\
& + \frac{C_{LNLe}^{\beta N\alpha\beta}}{\Lambda^2}\, \overline{u}(\ell_\alpha) P_R u(\ell_\beta)\,\, \overline{u}(\nu_\beta) P_R v(\overline{N}) - \frac{C_{LNLe}^{\alpha N\beta\beta}}{\Lambda^2}\, \overline{u}(\ell_\alpha) P_R v(\overline{N})\,\, \overline{u}(\nu_\beta)P_R u(\ell_\beta),\\
&\frac{1}{2}\sum |\mathcal{M}|^2 = \frac{2}{M_W^2\Lambda^4} \Big[ (k_2.k_3) (p.k_1)\Big\{ M_W^2 \left(4(C_{HNe}^{N\alpha})^2+(C_{LNLe}^{\beta N\alpha\beta})^2-C_{LNLe}^{\beta N\alpha\beta} C_{LNLe}^{\alpha N\beta\beta}\right)+8(C_{LNW}^{\alpha N})^2 m_{\ell_\beta}^2\Big\}\nonumber \\
& + 32 (k_2.k_3) (C_{LNW}^{\alpha N})^2 (k_1.k_3) \left((p.k_3)-m_{\ell_\beta}^2\right) + 2 C_{HNe}^{N\alpha} \left(2C_{LNLe}^{\beta N\alpha\beta}-C_{LNLe}^{\alpha N\beta\beta}\right) (k_2.k_3) M_W^2 m_{\ell_\alpha} m_{\ell_\beta}\nonumber \\
& + (p.k_2) (k_1.k_3) \left(C_{LNLe}^{\beta N\alpha\beta} C_{LNLe}^{\alpha N\beta\beta} M_W^2 + 8 (C_{LNW}^{\alpha N})^2 m_{\ell_\beta}^2\right) + (p.k_3) (k_1.k_2) \Big( M_W^2 C_{LNLe}^{\alpha N\beta\beta} (C_{LNLe}^{\alpha N\beta\beta}-C_{LNLe}^{\beta N\alpha\beta})\nonumber \\
& -4(C_{LNW}^{\alpha N})^2 m_{\ell_\beta}^2\Big) \Big].
\end{align}
3. $\ell_\beta^-(p)\to\ell_\alpha^-(k_1) \overline{\nu_\alpha}(k_2) N(k_3)$:
\begin{align}
&\mathcal{M}(\ell_\beta^-\to\ell_\alpha^-N\overline{\nu_\alpha}) = \frac{C_{HNe}^{N\beta}}{\Lambda^2}\, \overline{u}(N)\gamma^\mu P_R u(\ell_\beta)\,\, \overline{u}(\ell_\alpha)\gamma_\mu P_L v(\overline{\nu_\alpha}) - \frac{2C_{LNW}^{\beta N}}{M_W\Lambda^2}\, \overline{u}(N)\sigma^{\mu\nu} P_L u(\ell_\beta) \,\, \overline{u}(\ell_\alpha)\gamma_\nu P_L v(\overline{\nu_\alpha}) (p_{\ell_\beta}-p_{N})_\mu\nonumber \\
& + \frac{C_{LNLe}^{\alpha N\beta\alpha}}{\Lambda^2}\, \overline{u}(\ell_\alpha) P_L u(\ell_\beta)\,\, \overline{u}(N) P_L v(\overline{\nu_\alpha}) - \frac{C_{LNLe}^{\beta N\alpha\alpha}}{\Lambda^2}\, \overline{u}(N) P_L u(\ell_\beta)\,\, \overline{u}(\ell_\alpha) P_L v(\overline{\nu_\alpha}),\\
&\frac{1}{2}\sum |\mathcal{M}|^2 =\frac{2}{M_W^2\Lambda^4} \Big[ -8(C_{LNW}^{\beta N})^2 (k_1.k_2) (p.k_3)^2 + (k_2.k_3) (p.k_1) \Big\{ 8 (C_{LNW}^{\beta N})^2 (p.k_3) + M_W^2 \Big(4(C_{HNe}^{N\beta})^2 \nonumber \\
& + (C_{LNLe}^{\alpha N\beta\alpha})^2-C_{LNLe}^{\alpha N\beta\alpha}C_{LNLe}^{\beta N\alpha\alpha}\Big)+8(C_{LNW}^{\beta N})^2 m_{\ell_\beta}^2\Big\} + 8 (C_{LNW}^{\beta N})^2 (k_2.k_3) (k_1.k_3) (p.k_3-2m_{\ell_\beta}^2)\nonumber \\
& + 2 (k_2.k_3) m_{\ell_\alpha} m_{\ell_\beta} M_W^2 C_{HNe}^{N\beta} \left(2 C_{LNLe}^{\alpha N\beta\alpha}-C_{LNLe}^{\beta N\alpha\alpha}\right) + (p.k_3) (k_1.k_2)\Big\{M_W^2 C_{LNLe}^{\beta N\alpha\alpha}\left(C_{LNLe}^{\beta N\alpha\alpha}-C_{LNLe}^{\alpha N\beta\alpha}\right)\nonumber \\
& +4(C_{LNW}^{\beta N})^2 m_{\ell_\beta}^2\Big\} + 8 (C_{LNW}^{\beta N})^2 (p.k_3) (p.k_2) (p.k_1+k_1.k_3) + (p.k_2) (k_1.k_3) \Big(C_{LNLe}^{\alpha N\beta\alpha} C_{LNLe}^{\beta N\alpha\alpha} M_W^2\nonumber \\
& - 8 (C_{LNW}^{\beta N})^2 m_{\ell_\beta}^2\Big)\Big].
\end{align}
4. $\ell_\beta^-(p)\to\ell_\alpha^-(k_1) \overline{N}(k_2) \nu_i(k_3)$:
\begin{align}
&\mathcal{M}(\ell_\beta^-\to\ell_\alpha^-\overline{N}\nu_i) = \frac{C_{LNLe}^{iN\alpha\beta}}{\Lambda^2}\, \overline{u}(\ell_\alpha)P_R u(\ell_\beta)\,\, \overline{u}(\nu_i) P_R v(\overline{N}) - \frac{C_{LNLe}^{\alpha N i\beta}}{\Lambda^2} \, \overline{u}(\ell_\alpha)P_R v(\overline{N})\,\, \overline{u}(\nu_i) P_R u(\ell_\beta),\\
&\frac{1}{2}\sum |\mathcal{M}|^2 =\frac{2}{\Lambda^4}\Big[ C_{LNLe}^{iN\alpha\beta} \left(C_{LNLe}^{iN\alpha\beta}-C_{LNLe}^{\alpha N i\beta}\right) (p.k_1) (k_2.k_3) + C_{LNLe}^{iN\alpha\beta} C_{LNLe}^{\alpha Ni\beta} (p.k_2) (k_1.k_3) \nonumber \\
& + C_{LNLe}^{\alpha Ni\beta} \left(C_{LNLe}^{\alpha Ni\beta}-C_{LNLe}^{iN\alpha\beta}\right) (p.k_3) (k_1.k_2)\Big].
\end{align}
5. $\ell_\beta^-(p)\to\ell_\alpha^-(k_1) \overline{\nu_i}(k_2) N(k_3)$:
\begin{align}
&\mathcal{M}(\ell_\beta^-\to\ell_\alpha^-N \overline{\nu_i}) = \frac{C_{LNLe}^{iN\beta\alpha}}{\Lambda^2}\, \overline{u}(\ell_\alpha)P_L u(\ell_\beta)\,\, \overline{u}(N) P_L v(\overline{\nu_i}) - \frac{C_{LNLe}^{\beta N i\alpha}}{\Lambda^2} \, \overline{u}(N)P_L u(\ell_\beta)\,\, \overline{u}(\ell_\alpha) P_L v(\overline{\nu_i}),\\
&\frac{1}{2}\sum |\mathcal{M}|^2 =\frac{2}{\Lambda^4}\Big[ C_{LNLe}^{iN\alpha\beta} \left(C_{LNLe}^{iN\alpha\beta}-C_{LNLe}^{\alpha N i\beta}\right) (p.k_1) (k_2.k_3) + C_{LNLe}^{iN\alpha\beta} C_{LNLe}^{\alpha Ni\beta} (p.k_2) (k_1.k_3) \nonumber \\
& + C_{LNLe}^{\alpha Ni\beta} \left(C_{LNLe}^{\alpha Ni\beta}-C_{LNLe}^{iN\alpha\beta}\right) (p.k_3) (k_1.k_2)\Big].
\end{align}
\section{Analytical cross sections relevant for $N_{\rm eff}$ calculations}
\label{appen:Xsec}
\begin{table}[]
    \centering
    \begin{tabular}{||c|c||}
 \hline
 Involved Operators 
 & Vertices \\
 \hline\hline
 $\Lambda_{\rm LNB}$ & $\frac{v}{\sqrt{2}\Lambda_{LNB}^2}\,\overline{\nu}_L \sigma^{\mu \nu} N_R B_{\mu \nu}$ \\ 
 \hline
 $\Lambda_{\rm LNW}$ & $\frac{v}{\sqrt{2}\Lambda_{LNW}^2}\,\overline{\nu}_L \sigma^{\mu \nu} N_R W^3_{\mu \nu}$\\
 \hline
 $\Lambda_{\rm LNW}$ &  $\frac{v}{\sqrt{2}\Lambda_{LNW}^2}\,\overline{e}_L \sigma^{\mu \nu} N_R \left(W_{\mu \nu}^1 + i W_{\mu \nu}^2 \right)$\\
 \hline
 $\Lambda_{\rm HNe}$ & $- \frac{v^2 g_L}{\sqrt{2}\Lambda_{HNe}^2}\overline{N}_R \gamma^\mu e_R W_\mu^+$\\
 \hline
  $\Lambda_{\rm HN}$ & $ \frac{v^2 g_L}{2C_W \Lambda_{HN}^2}\overline{N}_R \gamma^\mu N_R Z_\mu$\\
 \hline
\end{tabular}
    \caption{Vertex factor along with the interacting fields has been shown for the relevant operators.}
    \label{tab:Vertex}
\end{table}
In the following section, we have given the analytical expressions for the cross sections for the Feynman diagrams that are responsible for thermalizing $N_R$ in the early Universe, as shown in Figure~\ref{fig:NRtherm}. 
{\small \begin{eqnarray}
\sigma_{LNB}(\overline{\nu_{L}}N_R \rightarrow e^+ e^-) &=& \frac{\alpha v^2 }{12 \Lambda_{LNB}^4} \left(2 C_W^2 + \frac{s^2 \left(2- 2 \cos{2\theta_W}  + 4 \cos{4\theta_W}\right) \sec{\theta_W}}{ 4 \left(s -  M_Z^2\right)^2}  + \frac{s \left(1 - 2 \cos{2 \theta_W}\right)}{\left(s-M_Z^2\right)}\right)
\end{eqnarray}}
{\small \begin{eqnarray}\nonumber
\sigma_{LNW}(\overline{\nu_{L}}N_R \rightarrow e^+ e^-)&=& \frac{\alpha v^2}{12 \Lambda_{LNW}^4} \left[ 2 S_W^2 + \frac{s^2 \left(2- 2 \cos{2\theta_W}  + 4 \cos{4\theta_W}\right) \csc{\theta_W}}{ 4 \left(s -  M_Z^2\right)^2}  + \frac{s \left(1 - 2 \cos{2 \theta_W}\right)}{\left(s-M_Z^2\right)} - \right. \\ 
&&\left. \frac{\csc{\theta_W}^2}{s} \left\{2s + (s+2M_W^2) \log{\left(\frac{M_W^2}{s+M_W^2}\right)} \right\}
\right]
\end{eqnarray}}
{\small \begin{eqnarray}
\sigma_{HN}(\overline{N_R} N_R \rightarrow e+ e^-) &=& \frac{\pi  \alpha^2  s v^4 
\left(\left(C_W^2-3 S_W^2\right)^2+1\right)}{24 \Lambda_{HN}^4 C_W^4 S_W^4 \left(M_Z^2-s\right)^2}
\end{eqnarray}}
{\small \begin{eqnarray}
\sigma_{HNe}(e^+ N_R \rightarrow \nu e^+) &=& \frac{\pi\alpha^2  v^4} {S_W^4 \Lambda_{HNe}^4  s^2\left (M_{W}^2 + 
     s \right)}  \left(2\left (M_{W}^2 + 
          s \right)^2\left (\log\left(\frac{M_{W}^2}{M_{W}^2 + s} \right) \right) + 
     s\left (\frac {2\left (M_{W}^4 + s^2 \right)} {M_{W}^2} + 
        3  s \right) \right) 
\end{eqnarray}}

{\small \begin{eqnarray}
\sigma_{LNB}(e^+ N_R\rightarrow e^+ \nu) &=& \frac {\alpha v^2} {8 s \Lambda_{LNB}^4} \Bigg[ {4 C_W^2 \left (2\left (2m_{A}^2 + 
         s \right)\log\left (\frac {m_{A}^2 + 
          s} {m_{A}^2} \right) - 4  s \right)} - \frac {  (-2 \cos  (2\theta_W) + \cos (4\theta_W) + 
      2)}{C_W^2}\times \nonumber \\
      && \left (
    ( 2M_Z^2+s) \log\left (\frac {M_Z^2} {M_Z^2 + 
          s} \right) + 2  s \right) + {2 (2\cos  (2 \theta_W) - 1)\left (\left (M_Z^2 + 
         s \right)\log\left (\frac {M_Z^2} {M_Z^2 + 
          s} \right) + s \right)} \Bigg]
\label{eq:sigLNB}
\end{eqnarray}}    

{\small \begin{eqnarray}
\sigma_{LNW}(e^+ N_R\rightarrow e^+ \nu) &=& \frac {\alpha v^2} {8 s \Lambda_{LNW}^4} \Bigg[ {8\left (M_{W}^2\log\left (\frac {M_{W}^2} {M_{W}^2 + s} \right) + s \right)} -\frac {4}{S_W^2}\left ( (2 M_{W}^2+s) \log\left(\frac {M_{W}^2} {M_{W}^2 + s} \right)+ 2  s \right) -\\ \nonumber 
&& \frac {  (-2 \cos  (2\theta_W) + \cos (4\theta_W) + 
      2)}{S_W^2} \left (
    ( 2M_Z^2+s) \log\left (\frac {M_Z^2} {M_Z^2 + 
          s} \right) + 2  s \right) + \\
          \nonumber &&  {2 (2\cos  (2 \theta_W) - 1)\left (\left (M_Z^2 + 
         s \right)\log\left (\frac {M_Z^2} {M_Z^2 + 
          s} \right) + s \right)} +  {4 S_W^2 \left (2\left (2m_{A}^2 + 
         s \right)\log\left (\frac {m_{A}^2 + 
          s} {m_{A}^2} \right) - 4  s \right)}  - \\ \nonumber && 
          \frac {4\left (\cot^2 \theta_W - 
      1 \right) }{ \left (M_W^2 + M_{Z}^2 + s \right)}\left (s\left (M_W^2 + M_Z^2 + 
        s \right) + M_{W}^2\left (M_W^2 + 
         s \right)\log\left (\frac {M_W^2} {M_W^2 + 
          s} \right) + M_{Z}^2\left (M_Z^2 + 
         s \right)\log\left (\frac {M_Z^2} {M_Z^2 + 
          s} \right) \right) \Bigg]
\label{eq:sigLNW}
\end{eqnarray}}
Here, $\theta_W$ is the Weinberg angle, $S_W$ and $C_W$ are $\sin{\theta_W}$ and $\cos{\theta_W}$ respectively. In the equation \eqref{eq:sigLNW}, we have used the thermal mass of the photon $m_A^2=\frac{e^2 T^{2}}{6}$.
\section{Derivation of the collision term $\mathcal{C}_{N_R}$}
\label{BEeq}
Here, we have shown the detailed derivation of the collision term $\mathcal{C}_{N_R}$ that we have used to solve the Eq.\,\ref{eq:BETNR}. 
\begin{eqnarray}
&&\mathcal{C}_{N_R}
=
\int \prod_{\alpha=1}^{4} d\Pi_{\alpha}\,
\left(2 \pi \right)^4 \delta^4\left({\bf p}_1 + {\bf p}_2
- {\bf p}_3 - {\bf p}_4 \right)
\overline{\left|\mathcal{M}\right|}^2_{{\bf N_R}+X\rightarrow Y+Z}
\left[f_{Y}(p_3,t) f_{Z}({p}_4,t) - 
f_{\bold{N_R}}({p}_1,t) f_{X}({p}_2,t)
\right]E_1\,,\nonumber \\
\label{eq:CNR1}
\end{eqnarray}
where $\bf{p}_{{}_i}$s ($i=1$ to $4$) are the
four momenta of $N_R$, $X$, $Y$ and $Z$ respectively and
the corresponding energy is denoted by $E_i$.
Considering the Maxwell-Boltzmann distribution for the equilibrium distribution function, one can write the out-of-equilibrium distribution function of a species
$i$ having energy $E_i$ and temperature $T_i$ as $f_i = \frac{n_i}{n_i^{eq}}\exp\left(-\frac{E_i}{T_i}\right)$ where $n_i$ is the number density of the particle $i$ and $n_i^{eq}$ is the equilibrium number density. As $X,\,Y,\,Z$ are part of the SM thermal plasma they have the same temperature. Therefore, equation \eqref{eq:CNR1} can be written as 
\begin{eqnarray}
\mathcal{C}_{N_R}
&=&
\int \prod_{\alpha=1}^{4} d\Pi_{\alpha}\,
\left(2 \pi \right)^4 \delta^4\left({\bf p}_1 + {\bf p}_2
- {\bf p}_3 - {\bf p}_4 \right) \nonumber \times \overline{\left|\mathcal{M}\right|}^2_{{\bf N_{R}}+X\rightarrow Y+Z} \\
&&
\left(\exp\left(-\frac{E_3+E_4}{T}\right)\dfrac{n_Y}{n^{eq}_Y} \dfrac{n_Z}{n^{eq}_Z}
- \exp\left(-\frac{E_1}{T_\bold{N_R}} - \frac{E_2}{T}\right)\dfrac{n_{\bold{N_R}}}{n^{eq}_{\bold{N_R}}}\dfrac{n_X}{n^{eq}_X}\right)\,E_1\,=
\mathcal{C}_1 - \mathcal{C}_2 \,,
\end{eqnarray}
where
\begin{eqnarray}
    \mathcal{C}_1 &=& \int \prod_{\alpha=1}^{4} d\Pi_{\alpha}\,
\left(2 \pi \right)^4 \delta^4\left({\bf p}_1 + {\bf p}_2
- {\bf p}_3 - {\bf p}_4 \right) \nonumber \times \overline{\left|\mathcal{M}\right|}^2_{{\bf N_R}+X\rightarrow Y+Z} \left(\exp\left(-\frac{E_3+E_4}{T}\right)\dfrac{n_Y}{n^{eq}_Y} \dfrac{n_Z}{n^{eq}_Z}\right)\,E_1\,\, \\
&=&
\int \prod_{\beta=1}^{2}\dfrac{g_{\beta}\,d^3 \vec{p}_\beta}{(2\pi)^3}
\Bigg\{\dfrac{1}{4\,E_1\,E_2} \int \prod_{\alpha=3}^{4} d\Pi_{\alpha}\,
\left(2 \pi \right)^4 \delta^4\left({\bf p}_1 + {\bf p}_2
- {\bf p}_3 - {\bf p}_4 \right)\overline{\left|\mathcal{M}\right|}^2_{{\bf N_R}+X\rightarrow Y+Z}
\Bigg\} E_1\,\exp\left(-\frac{E_1+E_2}{T}\right)
\dfrac{n_Y}{n^{eq}_Y} \dfrac{n_Z}{n^{eq}_Z}, \nonumber
\label{sigmadef} \\
&=& \int \prod_{\beta=1}^{2}\dfrac{g_{\beta}\,d^3
\vec{p}_\beta}{(2\pi)^3}\,
E_1\,\exp\left(-\frac{E_1+E_2}{T}\right)
\sigma_{{\bf N_R}+X\rightarrow Y+Z}\, v_{\bf rel}\,
\dfrac{n_Y}{n^{eq}_Y} \dfrac{n_Z}{n^{eq}_Z} \,.
\label{eq:CNR2}
\end{eqnarray}
Here $g_{\beta}$ is the internal degrees of freedom
of the species $\beta$ and $\overline{\left|\mathcal{M}\right|}^2_{{\bf N_R}+X\rightarrow Y+Z}$ is the matrix amplitude square for scattering
${\bf N_R}+X\rightarrow Y+Z$, and it is summed over the final states and averaged over the initial states.  
 By using the prescription given in \cite{Gondolo:1990dk}, we can replace the variables $E_1$, $E_2$, and $\theta$ by the three other variables as $E_{\pm} = E_1 \pm E_2$ and the Mandelstam variable\footnote{Here, we have neglected the masses of the initial and final state particles as they are much lighter than the temperature scale we are interested in.} $s=2E_1 E_2 - 2 p_1 p_2 \cos \theta$ and the volume element $d^3 \vec{p}_1 d^3 \vec{p}_2 = 2\pi^2 E_1 E_2 dE_+ dE_- ds$. In this case, the integration limits for the new variables can be written as $|E_-|\leq \sqrt{E_+^2 -s },\,E_+\geq \sqrt{s},\, s\geq 0$. The change in integration variables further simplifies the collision term
 for the forward scattering \eqref{eq:CNR2} as ,
\begin{eqnarray}
    \mathcal{C}_1
&=& \dfrac{n_Y}{n^{eq}_Y} \dfrac{n_Z}{n^{eq}_Z} \frac{g_{N_R} g_X \pi^2}{ (2\pi)^6} \int \sigma_{{\bf N_R}+X\rightarrow Y+Z}\,
{\bf F}(s)\,ds \int \exp\left(-\frac{E_+}{T}\right)dE_+\int dE_-\,
 (E_+ + E_-) \nonumber \\
&=& \dfrac{n_Y}{n^{eq}_Y} \dfrac{n_Z}{n^{eq}_Z} \frac{g_{N_R} g_X \pi^2  T}{(2\pi)^6} \int_{0}^{\infty} \sigma_{{\bf N_R}+X\rightarrow Y+Z}\,s^2 \,K_{2}\left(\frac{\sqrt{s}}{T} \right) ds\,, 
\end{eqnarray}
where the quantity ${\bf F} = E_1 E_2 v_{\bf rel}$
depends on $s$ only and in the centre of momentum
frame ${\bf F} = s/2$. 

Now, the collision term for the backward scattering process
can be expressed as 
\begin{eqnarray}
    \mathcal{C}_2 &=& \int \prod_{\alpha=1}^{4} d\Pi_{\alpha}\,
\left(2 \pi \right)^4 \delta^4\left({\bf p}_1 + {\bf p}_2
- {\bf p}_3 - {\bf p}_4 \right) \nonumber \times \overline{\left|\mathcal{M}\right|}^2_{{\bf N_R}+X\rightarrow Y+Z} \left(\exp\left(-\frac{E_1}{T_{N_R}} - \frac{E_2}{T}\right)\dfrac{n_{\bold{N_R}}}{n^{eq}_{\bold{N_R}}}\dfrac{n_X}{n^{eq}_X}\right)\,E_1\,\,\\
&=& \int \prod_{\beta=1}^{2}\dfrac{g_\beta\,d^3 \vec{p}_\beta}{(2\pi)^3}   E_1\,\exp\left(-\frac{E_1}{T_{N_R}} - \frac{E_2}{T}\right) \sigma_{{\bf N_R}+X\rightarrow Y+Z}\,v_{\bf rel}\,\dfrac{n_{\bold{N_R}}}{n^{eq}_{\bold{N_R}}}\dfrac{n_X}{n^{eq}_X}, \nonumber \\
&=&  \dfrac{n_{\bold{N_R}}}{n^{eq}_{\bold{N_R}}}\dfrac{n_X}{n^{eq}_X}\frac{g_{N_R} g_X \pi^2}{ (2\pi)^6} 
\int \sigma_{{\bf N_R}+X\rightarrow Y+Z} 
{\bf F}(s)\,ds \int \exp\left(-\frac{E_+}{T_\bold{+}}\right) dE_+\int dE_-\,\exp\left( - \frac{E_-}{T_-}\right)
(E_+ + E_-) \nonumber \\
&=& 
\dfrac{n_{\bold{N_R}}}{n^{eq}_{\bold{N_R}}}\dfrac{n_X}{n^{eq}_X}\frac{g_{N_R} g_X \pi^2 2 T_- }{ 2(2\pi)^6} \int_0^{\infty} \sigma_{{\bf N_R}+X\rightarrow Y+Z}\,s\,ds \int_{\sqrt{s}}^{\infty} \exp\left(-\frac{E_+}{T_\bold{+}}\right) \Bigg\{(E_+ +T_-) \sinh\left({\frac{\sqrt{E_+^2-s}}{T_-}}\right)- \nonumber \\
 &&~~~~~~~~~~~~~~~~~~~~~~~~
 ~~~~~~\sqrt{E_+^2 -s}\,\, \cosh\left({\frac{\sqrt{E_+^2-s}}{T_-}}\right)\Bigg\} dE_+,
 \label{eq:CNRnu}
\end{eqnarray}

where in the third line we combined $T_{N_R}$ and $T$ to form two new variables $T_\pm$ as $\frac{1}{T\pm} = \frac{1}{2}\left(\frac{1}{T_{N_R}} \pm \frac{1}{T} \right)$ to simplify the integrations. Finally, we have done the integration over $E_+$ and $s$ \eqref{eq:CNRnu} numerically. 

\bibliographystyle{utphys}
\bibliography{bibliography} 
\end{document}